\documentclass[aps,11pt,prd,groupedaddress,nofootinbib,notitlepage,eqsecnum,preprintnumbers]{revtex4-2}
\usepackage[utf8]{inputenc}
\usepackage{hyperref}
\usepackage{xcolor}
\usepackage{graphicx}
\usepackage{amsmath,amssymb}
\usepackage{bm}
\usepackage{comment}
\usepackage[shortlabels]{enumitem}
\usepackage{overpic}
\usepackage{ulem}
\usepackage{makecell}
\usepackage{tabularx}
\usepackage{booktabs}
\usepackage{diagbox}
\usepackage{pifont}
\usepackage{etoolbox} 
\usepackage{orcidlink}

\makeatletter
\newcommand{\customlabel}[2]{%
   \protected@write \@auxout {}{\string \newlabel {#1}{{#2}{\thepage}{#2}{#1}{}} }%
   \hypertarget{#1}{#2}
}
\makeatother

\begin{document}
\title{GW Backgrounds associated with PBHs}

\author{\textsc{Guillem Dom\`enech$^{a,b}$\orcidlink{0000-0003-2788-884X}}}
    \email{{guillem.domenech}@{itp.uni-hannover.de}}

\affiliation{$^a$Institute for Theoretical Physics, Leibniz University Hannover, Appelstraße 2, 30167 Hannover, Germany.}
\affiliation{$^b$ Max-Planck-Institut für Gravitationsphysik, Albert-Einstein-Institut, 30167 Hannover, Germany}

\begin{abstract}
PBH formation requires high-density regions in the (random) density field filling the primordial universe. While only the largest (and so rarest) overdensities collapse to form PBHs, the rest cause large anisotropic stresses, which are the source of GWs.  We provide an overview of the theoretical aspects of the GW backgrounds associated with PBHs from large primordial fluctuations. We consider GW backgrounds associated with PBH formation, PBH reheating and unresolved PBH binaries. We present several graphical summaries and illustrations for the busy reader. 
\end{abstract}

\maketitle

\section{Introduction \label{sec:introSGWB}}

Whenever there are fluctuations on top of the isotropic and homogeneous primordial universe, there is a secondary generation of GWs. Kenji Tomita was the first to write about this effect in the '70s in Ref.~\cite{Tomita}, saying that ``gravitational waves are induced by deformed density perturbations''. Our interpretation of Ref.~\cite{Tomita} is that by ``deformed density perturbations'' Tomita meant the anisotropic stress resulting from the presence of density perturbations. This will be clear later from Fig.~\ref{fig:illustrationTij}. This effect, which we will call induced GWs, was later studied in the '90s by Matarrese, Pantano and Saez \cite{Matarrese:1992rp,Matarrese:1993zf} and by  Matarrese, Mollerach and Bruni \cite{Matarrese:1997ay}. Ten years later, in 2006, we find a glimpse of the potential of induced GWs by Ananda, Clarkson and Wands \cite{Ananda:2006af}, where they considered an ``excess power in a single mode'' from the power spectrum measured by CMB observations (and also using a notation very similar to the current one). A very nice work by Baumann, Steinhardt, Takahashi and Ichiki \cite{Baumann:2007zm} hinted that induced GWs could be enhanced during a matter-dominated phase (see also Refs.~\cite{Assadullahi:2009nf,Alabidi:2013lya}). But, it was not until 2008 that Saito and Yokoyama \cite{Saito:2008jc,Saito:2009jt} made the connection between induced GWs and PBHs. Their idea was quickly followed by Bugaev and Klimai \cite{Bugaev:2009zh,Bugaev:2009kq,Bugaev:2010bb}. For more details on the early history of induced GWs and recent developments, we refer the reader to Ref.~\cite{Domenech:2021ztg} for a recent review on the topic by the author of this chapter. Other helpful reviews can be found in Refs.~\cite{Yuan:2021qgz,LISACosmologyWorkingGroup:2023njw,Domenech:2023jve,Domenech:2024cjn}.

Induced GWs are a crucial observable for any PBH scenario. As it will be evident by the end of this chapter, an induced GW background must be present if there is (or was) a significant fraction of PBHs in the universe. Though the opposite is not always true, the induced GW background might suggest, or strongly exclude, PBHs as a significant fraction of DM. This chapter will aim to introduce the main formulas and predictions for the GW backgrounds associated with PBHs, focusing on a qualitative and intuitive understanding of the physics behind such GWs and the interesting parameter space to be probed. The busy reader will find a graphical summary of the parameter space in Fig.~\ref{fig:illustrationformation+binary} (for GWs associated with PBH formation) and in Fig.~\ref{fig:illustrationPBHreheating} (for GWs associated with PBH reheating). Their corresponding physical picture of the generation of induced GWs is respectively illustrated in Figs.~\ref{fig:illustrationTij} and \ref{fig:illustrationPBHfluct}.

The chapter is organized as follows. We first explore in \S~\ref{sec:birdSGWB} general sources of GWs in cosmology. We then revisit in \S~\ref{sec:energySGWB} the definition of the energy density of GWs in cosmology, emphasizing potential ambiguities for induced GWs. In \S~\ref{sec:curvatureSGWB}, we present the general formulation of induced GWs and focus on the most studied. In \S~\ref{sec:reheatingSGWB}, we review the GWs from the PBH-dominated early universe. The GW background from PBH binaries is explored in \S~\ref{sec:binariesSGWB}. The chapter ends with final remarks in \S~\ref{sec:remarksSGWB}.

\section{Bird's eye view on secondary GWs \label{sec:birdSGWB}}

We start with a general discussion on GWs in cosmology, with the aim of understanding secondary GWs and how they differ from other sources. We use the notation ``secondary GWs'' to refer to any GW sourced at second order in cosmological perturbation theory using the linear solutions to cosmic fluctuations. Our basic framework, which we introduce in detail below, is a perfect fluid in a perturbed FLRW universe (for the basics of cosmology, we recommend Mukhanov's \cite{mukhanov2005physical} and Baumann's \cite{Baumann:2022mni} books). A perfect fluid could be an adiabatic perfect fluid with a given equation of state $w=P/\rho$, where $P$ and $\rho$, respectively, are the pressure and energy density of the fluid. But, it could also be a scalar field, say $\varphi$. These two cases are described with the energy-momentum tensor given by
\begin{align}\label{eq:Tmunu}
T_{\mu\nu}&=\left(\rho+P\right)u_\mu u_\nu+P g_{\mu\nu}\,,
\end{align}
where $g_{\mu\nu}$ is our spacetime metric and $u_\mu$ the normalized 4-velocity of the fluid.\footnote{The canonical scalar field case is recovered by identifying
\begin{align}\label{eq:umutovarphi}
u_\mu=\frac{\partial_\mu \varphi}{\sqrt{-\partial_\alpha\varphi\partial^\alpha\varphi}}\quad,\quad
\rho=-\frac{1}{2}\partial_\alpha\varphi\partial^\alpha\varphi+V(\varphi)\,,\quad
P=-\frac{1}{2}\partial_\alpha\varphi\partial^\alpha\varphi-V(\varphi)\,,
\end{align}
where $V(\varphi)$ is the potential of the scalar field.}
The adiabatic perfect fluid describes, e.g., a gas of ultra-relativistic particles when $w=1/3$. This is the situation we encounter in standard Big Bang cosmology. A gas of non-relativistic particles has $w\ll1$, which is the case of standard cold DM. On the other hand, the scalar field can be used to describe inflation, reheating scenarios, bubble nucleation and topological defects.

To get to the equations for GWs, we begin with Einstein Equations, namely $G_{\mu\nu}=T_{\mu\nu}$ with $G_{\mu\nu}$ the Einstein tensor, and perturb both the metric and the matter fields. For simplicity, we work in conformal time and in the shear-free slicing (also known as the Newton gauge) in which the line element reads
\begin{align}\label{eq:dsperturbedFLRW}
ds^2=a^2\left(-e^{2\Phi}d\tau^2+e^{2\Psi}\left(e^h\right)_{ij}dx^idx^j\right)\,,
\end{align}
where $a$ is the scale factor, $\tau$ is conformal time, $\Phi$ and $\Psi$ are the gravitational potentials, $h_{ij}$ are the transverse-traceless degrees of freedom, and we neglected the transverse degrees of freedom, i.e., vector fluctuations. Note that we took the exponential ansatz for later convenience when discussing subtleties of second-order perturbation theory. For the matter fields, we have that
\begin{align}\label{eq:perturbedmatter}
u_0=-a(1+\Psi)\quad,\quad u_i=av_i\quad,\quad \rho=\rho(t)+\delta\rho\quad,\quad P= P(t)+\delta P\,,
\end{align}
where now $\rho(t)$ and $P(t)$ are the background components, i.e. homogeneous in space, and $\delta\rho$ and $\delta P$ are the perturbations. Similarly, we take $\varphi\to\varphi(t)+\delta\varphi$ for the scalar field case.

%\begin{align}\label{eq:GWequation1}
%\left[G_{ij}^{(1)}\right]^{\rm TT}=h_{ij}''+2{\cal H} h_{ij}'-\Delta h_{ij}=\left[G_{ij}^{(2)}+T_{ij}^{(2)}\right]^{\rm TT}\,,
%\end{align}

With the perturbative expansion given by Eqs.~\eqref{eq:dsperturbedFLRW} and \eqref{eq:perturbedmatter}, we may write the GW equation as \cite{Domenech:2017ems,Domenech:2021ztg} 
\begin{align}\label{eq:GWs.fluid}
h_{ij}''+2{\cal H} h_{ij}'-\Delta h_{ij}=\left[4\partial_i\Phi\partial_j\Phi+2a^2(\rho+P)v_iv_j\right]^{\rm TT}\,,
\end{align}
where $'=d/d\tau$, ${\cal H}=a'/a$, $\Delta=\delta^{ij}\partial_i\partial_j$ and we expanded Einstein equations up to second order in perturbation theory. Also, ``TT'' refers to the transverse-traceless component of a given tensor (see, e.g., Ref.~\cite{Domenech:2021ztg} for the explicit form of the projector). There is no linear source to the GWs in Eq.~\eqref{eq:GWs.fluid} as a perfect fluid has no linear anisotropies. To derive the right-hand side of \eqref{eq:GWs.fluid}, we used that at linear order $\Psi=-\Phi$ as well as the properties of the transverse-traceless projector to write terms like $\Phi\partial_i\partial_j\Phi$ as $-\partial_i\Phi\partial_j\Phi$. It is interesting to note that in Eq.~\eqref{eq:GWs.fluid}, the term with $\Phi$ comes from $G_{ij}^{(2)}$ and the term with $v_i$ from $T_{ij}^{(2)}$, where the superscript ${(2)}$ means the expansion at second order. Thus, the term with $\Phi$ can be understood as the ``backreaction'' of the metric, while the term with $v_i$ is the standard matter source to GWs.
With the above equation, we may qualitatively discuss cosmic sources of GWs. 

Let us start with the case in which GWs are generated on scales much smaller than the Hubble radius (also called Hubble horizon) by a ``spectator'' field, i.e. the field responsible for the source of GWs is a subdominant component of the universe's energy density. This includes, e.g., nucleation of bubbles due to a first-order phase transition or the formation and evolution of a network of solitons, such as cosmic strings, domain walls, oscillons, etc. In this regime, also called the weak gravity limit, one may safely neglect the contribution from $\Phi$ in Eq.~\eqref{eq:GWs.fluid}. This is because, on subhorizon scales, the term containing $\Phi$ in Eq.~\eqref{eq:GWs.fluid} is suppressed with respect to the $v_i$ contribution by the ratio of energy densities between the spectator and the dominant field. Thus, GWs are sourced by the anisotropic stress resulting from the inhomogeneous velocity flows or scalar field profiles. For instance, in the case of bubble nucleation, one of the primary sources of GWs are the acoustic density waves following bubble collisions (though the collisions themselves and later turbulences also produce GWs; see Refs.~\cite{Caprini:2018mtu,Roshan:2024qnv} for recent reviews). For cosmic strings, it may be better to think of Eq.~\eqref{eq:GWs.fluid} in terms of the quadrupole formula (as for a local source, the volume integral of $T_{ij}$ is related to the second time derivative of the quadrupole). Then, a global cosmic string may be thought of as a ``vibrating cylinder'' with a length proportional to the Hubble radius. Its time-dependent quadrupole is then the source of GWs. There are, of course, other processes that lead to GW production, such as cosmic string loops and kinks. As the present discussion is very naive, and the above-mentioned process entirely non-linear,  we refer the interested reader again to Refs.~\cite{Caprini:2018mtu,Roshan:2024qnv} for recent reviews with more detailed discussions and references therein.

\begin{figure}
\includegraphics[width=\columnwidth]{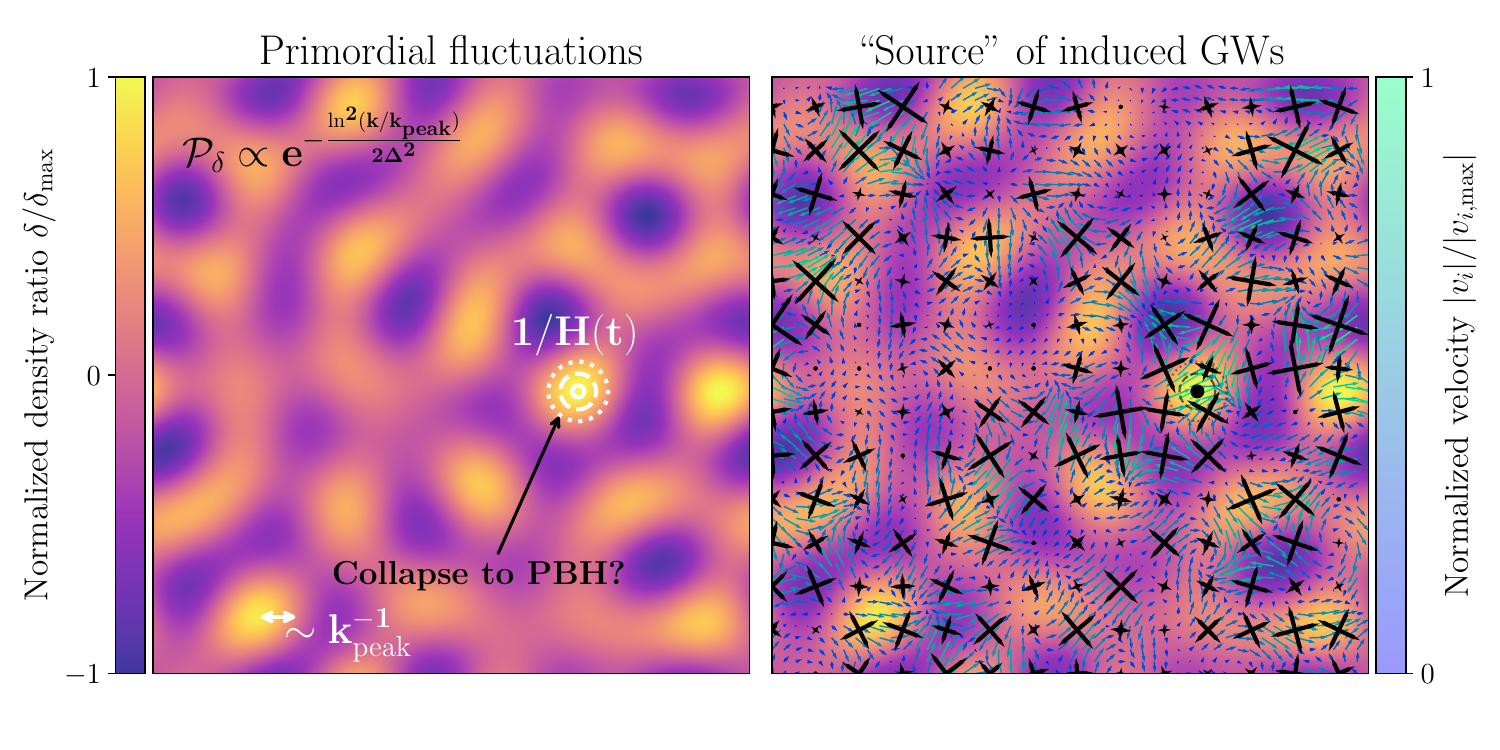}
\caption{Real space density field in 2D from Gaussian fluctuations with log-normal primordial spectrum with $\Delta=0.01$ and $k_{\rm peak}=0.005 L_{\rm box}^{-1}$ (i.e. the typical size of fluctuations is $20$ times smaller than the box), all in arbitrary units. On the left panel, we show the density ratio and illustrate that only the rarest and largest fluctuations collapse to form PBHs. On the right panel, we show the velocity gradients and the eigenvectors times the eigenvalues of the traceless component of the energy-momentum tensor. This exemplifies that induced GWs are generated by the mean fluctuations where gradients are largest.}\label{fig:illustrationTij} 
\end{figure}

A somewhat similar situation applies to secondary GWs associated with PBHs. But, in contrast with previous cases, secondary GWs are generated by the dominant fluid in the universe. And, the dominant fluid in the universe determines the \textit{curvature fluctuations}. Thus, we cannot generally neglect the term $\partial_i\Phi\partial_j\Phi$ in Eq.~\eqref{eq:GWs.fluid}. This is why one says that the GWs are induced by curvature fluctuations and are called induced GWs for short. We illustrate the physical picture of the generation of induced GWs in Fig.~\ref{fig:illustrationTij}. There, we considered a universe filled with random Gaussian superhorizon curvature fluctuations with a sharp log-normal power spectrum, peaked at $k_{\rm peak}$ and with a large enough amplitude. The left plot of Fig.~\ref{fig:illustrationTij} shows the 2D real space distribution of the density field, say $\delta$, or the curvature fluctuation. We see that only a few lumps (in this case, roughly one) have a large enough amplitude to eventually collapse and form a PBH once the Hubble horizon is approximately the size of the overdensity. The right plot of Fig.~\ref{fig:illustrationTij} illustrates with blue arrows the velocity flow (as being proportional to $\partial_i\delta$ or $\partial_i\Phi$) and with black crosses the eigenvectors of the traceless part of $T_{ij}$ (proportional to $v_iv_j$). Larger velocities have larger arrows and a lighter blue colour. The size of the crosses is proportional to the eigenvalues of the traceless part of $T_{ij}$.\footnote{We only consider the traceless part of $T_{ij}$ as in 2D there are no transverse-traceless degrees of freedom. Nevertheless, we expect that a similar situation applies to a more realistic 3D picture.} See how larger ``sources'' to the induced GWs are associated with larger velocity flows. We also see that while only one overdensity collapses to form a PBH, the rest source the induced GWs.

\begin{figure}
\includegraphics[width=\columnwidth]{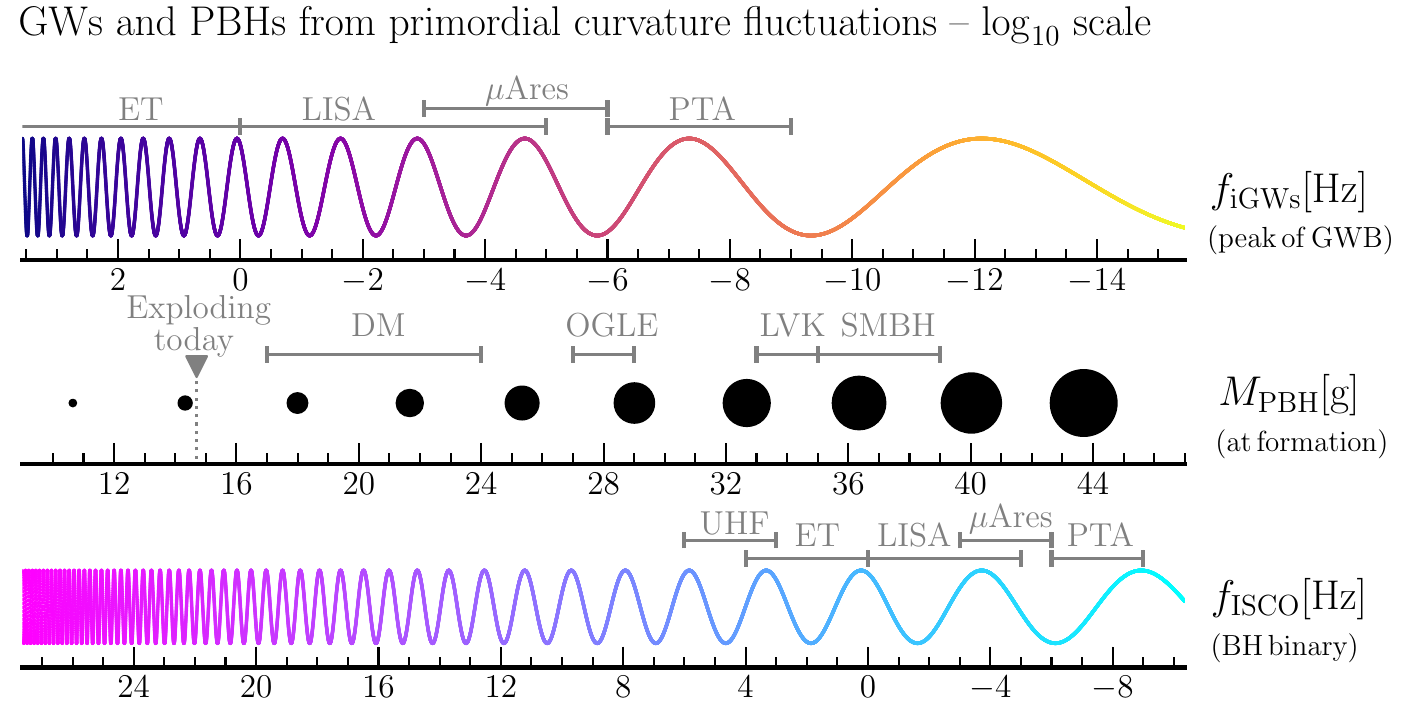}
\caption{Graphical summary of the relevant parameter space regarding GW frequency and PBH mass for the induced GWs (associated with PBH formation). In the top we show the frequency of induced GWs $\log_{10} f_{\rm iGWs}$ \eqref{eq:fiGWs}, in the middle we show PBH mass $\log_{10} M_{\rm PBH}$ and in the bottom we show the frequency of PBH mergers $\log_{10} f_{\rm ISCO}$  \eqref{eq:fisco}. They are all related to $M_{\rm PBH}$ by Eq.~\eqref{eq:PBHmass} through $k_{\rm peak}=a_{\rm f}H_{\rm f}$. We show the frequency range of current and future GW detectors such as Ultra-High-Frequency (UHF) GW detectors, ET, LISA, $\mu$-Ares and PTAs. We also draw the interesting PBH mass windows: PBH as all DM \cite{Green:2020jor,Carr:2020xqk}, planet mass PBH to explain OGLE candidate event \cite{Mroz:2017mvf,Niikura:2019kqi}, the merger of solar mass PBHs at LVK \cite{Bird:2016dcv,Sasaki:2016jop,Franciolini:2021tla} and PBH as the seeds of SMBH \cite{Carr:2018rid}. Note that the PTA window for induced GW overlaps with OGLE and LVK. \label{fig:illustrationformation+binary}}
\end{figure}

To put some units, we show in Fig.~\ref{fig:illustrationformation+binary} the frequency of the induced GWs $f_{\rm iGWs}$ (see Eq.~\eqref{eq:fiGWs}), the typical mass of the PBHs $M_{\rm PBH,f}$ and the frequency related to the Innermost Stable Circular Orbit (ISCO),  $f_{\rm ISCO}$, of the merger of an equal PBH binary (in the binary frame, see Eq.~\eqref{eq:fisco}). The redshift factor is the main difference between $f_{\rm iGW}$ and $f_{\rm ISCO}$. While induced GWs are generated at the same time PBHs form, the detected PBH binaries merge in the nearby universe. See how with GW detectors such as ET, LISA and PTAs, we can indirectly probe the existence of PBH with masses ranging from ${\cal O}(10^9\,{\rm g})$ to ${\cal O}(M_\odot)$ with induced GWs. In fact, there may already be hints of an induced GW background at nHz frequency from PTAs \cite{NANOGrav:2023hvm}. Interestingly, the same GW detectors can probe the mass range from ${\cal O}(M_\odot)$ to ${\cal O}(10^{9}M_\odot)$ from the merger of PBHs binaries. In the same figure we illustrate the most interesting PBH mass windows to explain the DM \cite{Carr:2020xqk}, the planet mass candidates events in OGLE \cite{Mroz:2017mvf,Niikura:2019kqi}, the BH mergers seen by the LVK collaboration \cite{Bird:2016dcv,Sasaki:2016jop,Franciolini:2021tla} and the seeds of supermassive black holes (SMBHs) \cite{Carr:2018rid}.

\section{Effective energy density of GWs in cosmology\label{sec:energySGWB}}

Before we dig into the details of the calculations of induced GWs, we briefly revisit the definition of the energy density of GWs in cosmology, as we will recursively use it to derive predictions. We will discuss important subtleties of the definition of energy density of the induced GWs at the end of this section. The derivation of the energy-momentum tensor can be found, e.g., in Maggiore's book \cite{Maggiore:1900zz}. A detailed discussion on the subtleties is given in Ref.~\cite{Domenech:2020xin}, and references therein.

\subsection{The pseudo Energy Momentum Tensor of GWs \label{subsec:EMT}}

To estimate the energy density carried by the GWs, we need to find the ``backreaction'' of \textit{linear} tensor fluctuations onto the background metric. To do that, we expand Einstein equations to second order (as we did in Eq.~\eqref{eq:GWs.fluid}), and compute some kind of average. This average could be a volume and/or a time average, inside which several GW wavelengths or oscillations fit. In the case of stochastic fields, the volume/time average is equivalent to the ensemble average (by ergodicity).  Expanding and taking the average, we formally find that
\begin{align}\label{eq:Gmunuexpansion2}
G_{\mu\nu}^{(0)}[a]\approx -\langle G^{(2)}_{\mu\nu}[a(\eta+h)]\rangle+T^{(0)}_{\mu\nu}\approx  t^{\rm GW}_{\mu\nu}+ T^{(0)}_{\mu\nu}\,,
\end{align}
where $\eta$ refers to the Minkowski metric $\eta_{\mu\nu}$, $h$ to $h_{ij}$ of Eq.~\eqref{eq:dsperturbedFLRW} and the brackets denote the chosen average. We moved the second order expansion $G^{(2)}_{\mu\nu}$ to the right-hand side so that we can interpret it as the energy-momentum tensor, which we call $t^{\rm GW}_{\mu\nu}$ in the last step. The contribution from linear fluctuations vanishes because of the linear equations of motion. After some algebra, one obtains that
\begin{equation}\label{eq:pseudotensor}
t_{\mu\nu}^{\rm GW}=\frac{1}{4}\left\langle\partial_\mu h^{ij}\partial_\nu h_{ij}-\frac{1}{2} g_{\mu\nu} \partial_\sigma h^{ij} \partial^\sigma h_{ij}\right\rangle\,.
\end{equation}
A fundamental assumption in deriving Eq.~\eqref{eq:pseudotensor} is that we only consider subhorizon tensor fluctuations $h_{ij}$. This gives us an unambiguous notion of a slowly evolving, large-scale background and high-frequency, small-scale, GWs. It is not clear how to apply a similar procedure when both scales are similar, e.g., for superhorizon tensor modes. Also, note that $t_{\mu\nu}^{\rm GW}$ is called a pseudotensor as it only transforms as a tensor at the linear level in perturbation theory. It is a priory not well-defined at higher orders.

The energy density of the GWs is given by the $00$ component of Eq.~\eqref{eq:pseudotensor}, namely $a^2\rho_{\rm GW}=t_{00}^{\rm GW}$ in conformal time. Since $h_{ij}$ is a random field, we can compute the energy density in terms of its power spectrum. Working with Fourier modes,\footnote{Our convention is that 
\begin{align}\label{eq:hijfourier}
h_{ij}=\frac{1}{(2\pi)^3}\sum_\lambda\int d^3k \,h_{\mathbf{k},\lambda}(\tau)\epsilon_{ij}^\lambda(\mathbf{k})\, e^{i\mathbf{k}\cdot\mathbf{x}}\,,
\end{align}
where polarization tensors are normalized, i.e.
$
(\epsilon_{ij}^\lambda(\mathbf{k}))^*\epsilon^{ij,\lambda'}(\mathbf{k})=\delta^{\lambda\lambda'}
$.
} the GW energy density fraction per $\ln k$, also referred to as the spectral density, is given by
\begin{equation}\label{eq:spectraldensity}
\Omega_{\rm GW}(k,\tau)=\frac{1}{\rho_{\rm tot}}\frac{d \rho_{\rm GW}}{d\ln k}=\frac{k^2}{12{\cal H}^2}\sum_\lambda\overline{{\cal P}_{h,\lambda}(k,\tau)}\,,
\end{equation}
where ${\cal P}_{h,\lambda}(k,\tau)$ is the dimensionless strain power spectrum, defined by
\begin{align}\label{eq:spectrumh}
\langle h_{\mathbf{k},\lambda}(\tau)h_{\mathbf{k}',\lambda'}(\tau)\rangle=(2\pi)^3 \delta(\mathbf{k}+\mathbf{k}')\delta_{\lambda\lambda'}\tfrac{2\pi^2}{k^3}{\cal P}_{h,\lambda}(k,\tau)\,,
\end{align}
and we took an additional oscillation (time) average denoted with an overline, as the same average is done when looking for GW backgrounds. It should be noted that in deriving Eq.~\eqref{eq:spectraldensity} we used that for subhorizon GWs $k\gg {\cal H}$ and that $h'_{\mathbf{k},\lambda}\sim k h_{\mathbf{k},\lambda}$. Eq.~\eqref{eq:spectraldensity} is the quantity we will use to derive and compare our predictions with observations. The energy density of GWs is related to the strain noise variance.

\subsection{Subtleties of induced GWs \label{subsec:subtleties}}

Before ending this section, let us point out the fundamental ambiguity in the definition \eqref{eq:pseudotensor} of the energy density of GWs. We can convince ourselves that the definition is ambiguous by checking whether $\rho_{\rm GW}$ is invariant under a shift of the conformal time coordinate, e.g. $\bar\tau=\tau+T(\tau,x^i)$,
where $T$ is an infinitesimal gauge parameter. With some effort, one finds that the tensor perturbations defined as in Eq.~\eqref{eq:dsperturbedFLRW} transform according to \cite{Domenech:2017ems}
\begin{align}\label{eq:gaugetransformationhij}
\bar h_{ij}=h_{ij}+\widehat{TT}^{kl}_{ij}\left[T \,h'_{kl}+\partial_kT\partial_l T\right]+...\,,
\end{align}
where ``$...$'' means higher order combinations in $T$ and $h_{ij}$ and we neglected vector perturbations. It turns out that what we called tensor perturbations $\bar h_{ij}$, in the barred time coordinate, are a mixture of tensor and scalar perturbations in the unbarred time coordinate. In other words, the definition of tensor perturbations beyond the linear order depends on the time slicing. The energy density of GWs as defined by Eq.~\eqref{eq:pseudotensor} is certainly not invariant at second order under the time shift \eqref{eq:gaugetransformationhij}.

It is not so surprising that Eq.~\eqref{eq:pseudotensor} becomes problematic at the second order. First, we considered only linear perturbations, and second, there is no local notion of energy in gravity, as we can always go to a local inertial (Minkowski) frame. Although this is a general problem, it is not so problematic for GW sources in the weak gravity regime, such as cosmic strings and phase transitions (where secondary GWs from $\Phi$ are already neglected), unless one deliberately chooses a strange time coordinate. But, it may be problematic for induced GWs since they are a second-order effect from $\Phi$. For instance, considering that the unbarred time corresponds to the time in the Newton gauge, we have that $T=T(\Phi,\tau)$, with the exact functional form depending on the chosen barred time slicing. Since $h^{\rm iGW}_{ij}\propto \partial_i\Phi\partial_j\Phi$ all the terms in the right-hand side of Eq.~\eqref{eq:gaugetransformationhij} can be in principle comparable. This issue was first pointed out by Ref.~\cite{Hwang:2017oxa} for a matter-dominated universe, in Ref.~\cite{Gong:2019mui} in radiation domination and in general in Ref.~\cite{Tomikawa:2019tvi}. See also Ref.~\cite{Sipp:2022kmb}. It was then argued in Refs.~\cite{DeLuca:2019ufz,Inomata:2019yww} that in radiation domination, the synchronous (which should describe the frame of a GW detector) and Newton gauges give the same results at late times. Then, in Ref.~\cite{Domenech:2020xin}, it was shown that such approximate gauge invariance applies to other gauges and for any cosmological background (except matter domination; see below). 

The crucial point to realize, which provides a partial resolution, is that we measure GWs far from the source \cite{Domenech:2020xin}. The same is required, e.g., to define the GWs from binaries. In cosmology, ``far'' from the source means that there is effectively no more production of GWs at some point in time. This is precisely the situation for induced GWs. The curvature perturbation $\Phi$ quickly decays once it enters the Hubble horizon, rendering the energy density of induced GWs practically independent of the time slicing.\footnote{By almost independent, we mean that the prediction for the energy density of induced GWs is independent of the time slicing up to corrections suppressed by $(k/{\cal H})^2$ (a ratio which evaluated today is tiny) for a restricted class of time slicings.} For instance, the calculations in the Newton gauge, constant Hubble gauge, flat gauge and synchronous gauge (after proper gauge fixing) all yield the same results. The exceptions to this statement are $(i)$ comoving slices and $(ii)$ a dust-dominated universe. In exception $(i)$, one can argue that such time slices are suitable for superhorizon but not for subhorizon physics. For example, on small scales, the velocity flow can become quite non-trivial and choosing a time slice that is comoving with the fluid everywhere is undoubtedly a convoluted choice. Thus, it is not surprising that the comoving slicing might contain gauge artefacts that look like GWs. In exception $(ii)$, the curvature perturbation is constant on all scales, so there is a constant source to the induced GWs. In this case, one must follow the GW production until the end of the dust-dominated era to have an unambiguous definition of $\Omega_{\rm GW}$. The resolution to the ambiguity lies in a systematic study of the strain measured by a GW detector at second order in cosmological perturbation theory. More details and references on the gauge issue can be found in Ref.~\cite{Domenech:2021ztg}.

\section{GWs from primordial curvature fluctuations  \label{sec:curvatureSGWB}}

After the general discussions in Secs.~\ref{sec:birdSGWB} and \ref{sec:energySGWB}, we focus on the mathematical formulation of induced GWs. We first take a general approach but quickly turn to the most studied case in Sec.~\ref{subsec:kernel}. Extensions of the formalism will be discussed in Sec.~\ref{sec:extensionsSGWB}. Details of the main calculations can be found in Refs.~\cite{Kohri:2018awv,Domenech:2019quo,Domenech:2021ztg}.

\subsection{General formulation\label{subsec:generalformulation}}

Consider that, after inflation, there are some initial curvature fluctuations $\Phi$ in the universe. Since induced GWs are sourced by the dominant fluid in the universe, the scalar component of velocity fluctuations in Eqs.~\eqref{eq:GWs.fluid} is also related to the curvature perturbation via Einstein equations (the $0i$ component to be precise). For this reason, we can take a general approach to the source term in Eqs.~\eqref{eq:GWs.fluid} by saying that, in Fourier modes,\footnote{The Fourier expansion for $\Phi$ follows the same convention as Eq.~\eqref{eq:hijfourier} for $h_{ij}$.} the source term can be split into a primordial value of the curvature perturbation $\Phi_{\mathbf{k}}^{\rm prim}$ times a time-dependent transfer function. A general equation for the induced tensor modes then reads
\begin{equation}\label{eq:eominducedGWfourier}
h''_{\mathbf{k},\lambda}+2{\cal H} h'_{\mathbf{k},\lambda}+k^2 h_{\mathbf{k},\lambda}=4\int_{} \frac{d^3q}{(2\pi)^3} e_\lambda^{ij}(k)q_iq_j f(\tau, q,|\mathbf{k-q}|)\Phi^{\rm prim}_{\mathbf{q}}\Phi^{\rm prim}_{|\mathbf{k-q}|}\,,
\end{equation}
where $f(\tau, q,|\mathbf{k-q}|)$ is the transfer function of the source. We will discuss particular cases later. We can also write down a formal integral solution to Eq.~\eqref{eq:eominducedGWfourier} by means of the Green's function method, which yields
\begin{align}\label{eq:greensolution}
h_{\mathbf{k},\lambda}(\tau)=\frac{4}{k^2}\int_{} \frac{d^3q}{(2\pi)^3}e_\lambda^{ij}(\mathbf{k})q_iq_jI(\tau,q,|\mathbf{k}-\mathbf{q}|) \Phi^{\rm prim}_{\mathbf{q}}\Phi^{\rm prim}_{|\mathbf{k-q}|} \,,
\end{align}
where we defined for compactness
\begin{align}\label{eq:Idefinition}
I(\tau,q,|\mathbf{k}-\mathbf{q}|)=k^2\int_0^\tau\!\!\! d\tau'\,G_h(\tau,\tau')f(\tau',q,|\mathbf{k}-\mathbf{q}|)\,.
\end{align}
$I(\tau,q,|\mathbf{k}-\mathbf{q}|)$ is often called the kernel of induced GWs, and $G_h$ in Eq.~\eqref{eq:Idefinition} is the Green's function for tensor modes which in radiation domination reads $kG_h(\tau,\tau')=\frac{a(\tau)}{a(\tau')}\sin(k(\tau-\tau'))$. The factors $k^2$ in Eqs.~\eqref{eq:greensolution} and \eqref{eq:Idefinition} are introduced for later convenience, as in this way, $I$ is dimensionless. 

With our solution, Eq.~\eqref{eq:greensolution}, we can compute the oscillation average of the dimensionless power spectrum that enters in the spectral density of GWs, Eq.~\eqref{eq:spectraldensity}. After some calculations, we obtain that
\begin{align}\label{eq:h2pt}
\overline{{\cal P}_{h,\lambda}(k,\tau)}=&\,\frac{4k}{\pi}\int_{} \frac{d^3q}{(2\pi)^3} \int_{}\frac{d^3q'}{(2\pi)^3}\overline{I(\tau,q,|\mathbf{k}-\mathbf{q}|)I(\tau,q',|\mathbf{k}-\mathbf{q}'|)}\nonumber\\[5pt]& \times e_\lambda^{ij}(\mathbf{k})q_iq_je_{\lambda}^{ij}(-\mathbf{k})q'_iq'_j
\langle\Phi^{\rm prim}_q\Phi^{\rm prim}_{|\mathbf{k}-\mathbf{q}|}\Phi^{\rm prim}_{q'}\Phi^{\rm prim}_{|\mathbf{k}-\mathbf{q}'|}\rangle'\,,
\end{align}
where we assumed that $\langle h_\lambda(k,\tau)h_{\lambda'}(k',\tau)\rangle$ follows the form of Eq.~\eqref{eq:spectrumh} so that we may subtract the factor $(2\pi)^3\delta_{\lambda\lambda'}\delta(\mathbf{k}+\mathbf{k}')$. We denote the remaining contribution to the 4-point function of $\Phi$ with a prime. Note that the oscillation average in Eq.~\eqref{eq:h2pt} must be done once the tensor modes are deep inside the Hubble horizon, i.e. for $k\tau\gg 1$. We restrict now to the case of Gaussian primordial fluctuations (which is often a good approximation), where the 4-point function is determined solely by the 2-point function. In this case, we find that
\begin{align}\label{eq:Phgaussian}
&\overline{{\cal P}_h(k,\tau)}=\sum_\lambda\overline{{\cal P}_{h,\lambda}(k,\tau)} \nonumber\\[5pt]&\quad=8\int_0^\infty\!\!\!dv\int_{|1-v|}^{1+v}\!\!\!du\left(\frac{4v^2-(1-u^2+v^2)^2}{4uv}\right)^2\overline{I^2(\tau,k,v,u)}{{\cal P}_{\Phi}(ku)}{{\cal P}_{\Phi}(kv)}\,,
\end{align}
where we introduced $v={q}/{k}$ and $u={|\mathbf{k}-\mathbf{q}|}/{k}$ for compactness.

It is worth noting that, in general, induced GWs are sourced by the typical curvature fluctuations, that is, $\langle\Phi^2\rangle^{1/2}$. In other words, the naive expectation is that $\langle\Phi^4\rangle\sim {\cal O}(\langle\Phi^2\rangle^2)$, unless the probability distribution function (PDF) of $\Phi$ is highly skewed (i.e. something very far from the Gaussian quantum fluctuations expected from inflation). In contrast, PBHs form from rare curvature fluctuations, which are determined by the tail of the PDF. Thus, while PBHs are exponentially sensitive to the root mean square of curvature fluctuations, induced GWs are proportional to the fourth power. From this, we conclude that the detection of induced GWs does not generally imply a significant fraction of PBHs in the universe. However, a substantial fraction of PBHs must have a detectable induced GW signal.

\subsection{Semi-analytical formula in radiation domination \label{subsec:kernel}}

Let us now revisit the most considered case, primordial adiabatic (curvature) fluctuations in a radiation-dominated universe. The first semi-analytical formula was derived in Refs.~\cite{Espinosa:2018eve,Kohri:2018awv} and was generalized in Refs.~\cite{Domenech:2019quo,Domenech:2020kqm}. We will discuss more possibilities in Sec.~\ref{sec:extensionsSGWB}. In the case that one fluid dominates the universe, the $0i$ component of Einstein's equations yields
\begin{align}\label{eq:vandPhi}
\Phi'+{\cal H}\Phi=-\frac{1}{2}a^2(\rho+P)v\,,
\end{align}
where $v$ is defined through $v_i=\partial_i v$. Additionally, the curvature fluctuations evolve according to
\begin{align}\label{eq:phigeneralsolutionRD}
\Phi(q\tau)=\Phi^{\rm prim}_{\mathbf{q}}\,\frac{3 j_1(c_sq\tau)}{c_sq\tau}\,,
\end{align}
where $j_1$ is the spherical Bessel function of order $1$.

Inserting the relation \eqref{eq:vandPhi} between $v$ and $\Phi$ in Eq.~\eqref{eq:GWs.fluid}, and using the linear solution Eq.~\eqref{eq:phigeneralsolutionRD}, we find the explicit form of $f(\tau,q,|\mathbf{k}-\mathbf{q}|)$ in Eq.~\eqref{eq:eominducedGWfourier}. The transfer function $f(\tau,q,|\mathbf{k}-\mathbf{q}|)$ is a product of sines and cosines, and the integral $I(\tau,q,|\mathbf{k}-\mathbf{q}|)$ \eqref{eq:Idefinition} can be carried out analytically. Then, combining Eq.~\eqref{eq:spectraldensity} and Eq.~\eqref{eq:Phgaussian}, we find a compact form for the GWs induced during standard radiation domination (that is $c_s^2=w=1/3$), namely
\begin{align}\label{eq:PhgaussianRD}
\Omega_{\rm GW,*}=\int_0^\infty dv\int_{|1-v|}^{1+v}du\,{\cal T}_{RD}(u,v){{\cal P}_{\cal R}(ku)}{{\cal P}_{\cal R}(kv)}\,,
\end{align}
where we used Eq.~\eqref{eq:phitoR} and defined
\begin{align}\label{eq:TRD}
{\cal T}_{RD}(u,v)=&\left(\frac{4v^2-(1-u^2+v^2)^2}{4u^2v^2}\right)^2\nonumber\\&\times y^2
\left\{\frac{\pi^2}{4}y^2\Theta[u+v-c_s^{-1}]
+\left(1-\frac{1}{2}y \ln\left|\frac{1+y}{1-y}\right|\right)^2\right\}\,,
\end{align}
and
\begin{align}\label{eq:y}
y=\frac{u^2+v^2-c_s^{-2}}{2uv}\,.
\end{align}
The meaning of the subscript “*” will be explained in the next paragraph.
Note that to obtain Eq.~\eqref{eq:TRD}, we took the limit $k\tau\to\infty$ of the integral \eqref{eq:Idefinition}. This is a very good approximation since we are interested in the GWs generated in the very early universe. So, any correction scales at least as ${\cal O}(1/(k\tau_{\rm eq}))$ where $\tau_{\rm eq}$ is the time of matter-radiation equality. Recall that the comoving scale that entered the Hubble horizon at that time is $k_{\rm eq}\sim 0.01 \,{\rm Mpc}^{-1}$, and we are interested in scales below a parsec. For the general form of the kernel for finite $k\tau$ see Ref.~\cite{Kohri:2018awv}. We also used that
\begin{align}\label{eq:phitoR}
	\Phi = \frac{3(1+w)}{5+3w}{\cal R}= \frac{2}{3}{\cal R}\,,
\end{align}
since predictions from inflation are given in terms of the curvature perturbation on comoving slices ${\cal R}$.
We note that the variable $y$, defined in Eq.~\eqref{eq:y}, is far from arbitrary. It is actually related to the area of the triangle composed by $1$, $c_sv$ and $c_su$ (or $k$, $c_sq$ and $c_s|\mathbf{k}-\mathbf{q}|$ in the original variables) and it naturally appears when integrating the product of three Bessel functions. In the end, it is a very useful variable for the extension to general constant $w$ and $c_s$. In the integration range of Eq.~\eqref{eq:Phgaussian}, we have that $y$ is bounded by $-\infty<y<1$.

Strictly speaking, instead of taking the limit $k\tau\to \infty$, one should integrate the kernel \eqref{eq:Idefinition} until today, that is at $\tau=\tau_0$. However, for the GW frequencies of interest, it is unnecessary to do the integrals across all the radiation, matter and dark energy domination phases. We may use the fact that GWs, once inside enough the Hubble horizon, behave as radiation. From there on, we can track the energy density ratio of GWs by comparing it to the one of standard radiation. Thus, if we call $\tau_*$ the time at which a given GW frequency starts behaving as radiation, we can write \cite{Inomata:2016rbd,Domenech:2021ztg}
\begin{align}\label{eq:GWstoday}.
\Omega_{\rm GW,0}h^2&=1.62\times 10^{-5}\left(\frac{\Omega_{\rm rad,0}h^2}{4.18\times 10^{-5}}\right)\left(\frac{g_{\rho}(T_*)}{106.75}\right)\left(\frac{g_{s}(T_*)}{106.75}\right)^{-4/3}\Omega_{\rm GW,*}\,,
\end{align}
where the subscript ``$*$'' means evaluation at $\tau=\tau_*$, $\Omega_{\rm rad,0}$ is energy density ratio of radiation today \cite{Planck:2018vyg} and $h=H_0/(100\,{\rm km/s/Mpc})$.
$g_{\rho}(T_*)$ and $g_{s}(T_*)$ are the effective degrees of freedom in the energy density and entropy of the standard radiation fluid. Thus, our goal is to estimate $\Omega_{\rm GW,*}$ and use Eq.~\eqref{eq:GWstoday} to find the amplitude of the induced GW background today. In passing, using that PBHs form with a mass given by
\begin{align}\label{eq:PBHmass}
M_{\rm PBH,f}=4\pi\gamma M_{\rm pl}^2/H_{\rm f}\,,
\end{align} 
where $\gamma\sim 0.2$ is an effective parameter (see, e.g., Ref.~\cite{Sasaki:2018dmp}), and that $k_{\rm peak}=a_{\rm f}H_{\rm f}$ where ``f'' refers to PBH formation, we can write $f_{\rm iGWs}$ of Fig.~\eqref{fig:illustrationformation+binary} in terms of $M_{\rm PBH,f}$ as 
\begin{align}\label{eq:fiGWs}
f_{\rm iGWs}\approx 1\,{\rm Hz}\left(\frac{M_{\rm PBH,f}}{10^{16}\rm g}\right)^{-1/2}\left(\frac{\gamma}{0.2}\right)^{1/2}\left(\frac{g_\rho(T_{*})}{106.75}\right)^{1/4}\left(\frac{  g_{s}(T_{*}) }{106.75}\right)^{-1/3}\,.
\end{align}

\subsection{Typical features and general properties \label{subsec:properties}}

Let us derive some useful and general properties of the induced GW spectrum from Eq.~\eqref{eq:PhgaussianRD}. This will provide us with intuition about the expected shape of the induced GW spectrum from a given primordial spectrum of curvature fluctuations, assuming that the primordial spectrum is peaked. We show a few examples of typical induced GW spectra in Fig.~\ref{fig:examplePBHreheating}.

\begin{figure}
\includegraphics[width=\columnwidth]{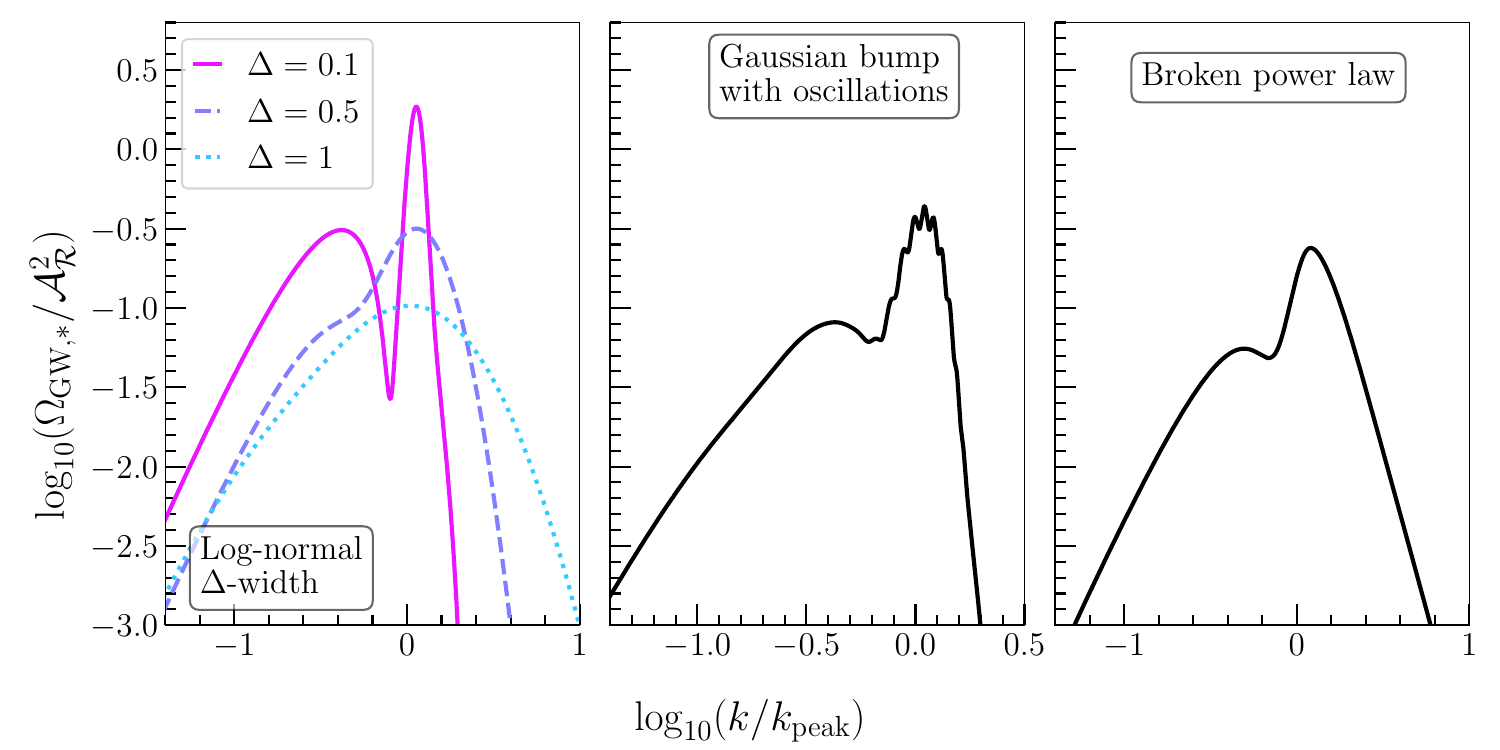}
\caption{Induced GW spectra in radiation domination computed numerically using SIGWfast \cite{Witkowski:2022mtg}. On the left panel, we show the induced GWs from a log-normal peak for various values of the logarithmic width $\Delta$. See how as the width increases the resonant peak is smoothed out. On the middle panel, we show the induced GW from a Gaussian bump with ${\cal O}(1)$ oscillations in the primordial spectrum from Ref.~\cite{Fumagalli:2020nvq}. Note how the oscillations in the induced GW spectrum are only ${\cal O}(10\%)$ modulations at the peak. On the right panel, we show the induced GW spectrum from a broken power-law primordial spectrum with $k^4$ in the IR and $k^{-5}$ in the UV. See how for all relatively sharp peaks, the induced GW spectrum has a similar shape. The IR tail of the induced GW spectrum in all cases goes as $k^3\ln^2 k$ (except for the very sharp $\Delta$ case, in magenta, which initially is $k^2\ln^2 k$).\label{fig:examplespectra}}
\end{figure}

The first thing one may notice from the averaged square kernel Eq.~\eqref{eq:TRD} is that it contains a logarithm that diverges at $y=\pm 1$. The value $y=-1$ occurs when $u+v=c_s^{-1}$. The origin of such divergence is a resonant production of induced tensor modes. In fact, Eq.~\eqref{eq:GWs.fluid} is nothing but a damped harmonic oscillator with a periodic force proportional to $\Phi^2$. So there is a resonance for $k=c_s(q+|\mathbf{k}-\mathbf{q}|)$, where $k$ is the tensor and $q$ the scalar wavenumbers.\footnote{This intuition is only valid if the source in Fourier space \eqref{eq:eominducedGWfourier} has a well-defined frequency. This only occurs when fluctuations of $\Phi$ have a typical scale, or in other words, the power spectrum of $\Phi$ is sufficiently peaked. From the point of view of the integral \eqref{eq:Phgaussian}, the exact resonance is a zero-measure point. So if the primordial spectrum is broad, the resonance is smeared out.} Thus, for a peaked enough spectra, the resonance lies at
\begin{align}\label{eq:kresonance}
k_{\rm res}=2c_sk_{\rm peak}.
\end{align}
The resonant peak is clearly visible in Fig.~\ref{fig:examplespectra}, where we show the induced GW spectra of a log-normal primordial spectrum with logarithmic with $\Delta$, a Gaussian bump with oscillations and a broken power-law spectrum, respectively from left to right. On the left plot, we also plot the induced GW spectrum for various values of the width. See how as one increases $\Delta$ the resonant peak disappears. For analytical approximations in the case of the log-normal primordial spectrum see Ref.~\cite{Pi:2020otn}. For the Gaussian bump with oscillations see Refs.~\cite{Fumagalli:2020nvq,Braglia:2020taf,Fumagalli:2021cel,Witkowski:2021raz}. And, for the broken power-law see Refs.~\cite{Atal:2021jyo,Xu:2019bdp,Liu:2020oqe}.

The value $y=1$ corresponds to the limit $u\sim v \gg 1$, which is related to long wavelength tensor modes $k\ll q$. Assuming that ${\cal P}_{\cal R}(k)\propto {\cal A}_{\cal R}(k/k_{\rm peak})^n$ (with appropriate cut-off) the very low frequency, or far Infra-Red (IR), tail reads \cite{Atal:2021jyo}
\begin{align}\label{eq:IRtail}
\Omega^{\rm IR}_{\rm GW,*}(k\ll k_{\rm peak})\propto
\begin{cases}
{\cal P}_{\cal R}^2(k) & {\rm for}\quad 0<n<3/2\\[3pt]
\left(\frac{k}{k_{\rm peak}}\right)^3\ln^2\left(\frac{k}{k_{\rm peak}}\right)& {\rm for}\quad n>3/2
\end{cases}
\,.
\end{align}
The first case of Eq.~\eqref{eq:IRtail} is also valid for broad enough peaks.\footnote{Looking at both Eqs.~\eqref{eq:IRtail} and \eqref{eq:UVtail} we see that for broad enough spectra, a good estimate to the induced GW spectrum is given by $\Omega_{\rm GW,*}(k)\sim
{\cal P}_{\cal R}^2(k)$. 
In the limit of a flat spectrum (that is ${\cal P}_{\cal R}\approx {\rm constant}$), one numerically finds that \cite{Kohri:2018awv}
\begin{align}
\label{eq:scaleinvariant}
\Omega^{\rm flat}_{\rm GW,*}\approx 0.82 \,{\cal P}_{\cal R}^2\,.
\end{align}} 
This is for example the case of a broad log-normal for scales not too far from the peak. For scales very far from the peak, one has a clear exponential decay, and one recovers the universal $k^3$ scaling. The second case corresponds to sharp enough peaks. Note that for the case of a Dirac delta primordial spectrum (corresponding to the $\Delta\to0$ limit of the log-normal), or a very sharp peak, the IR tail goes as $k^2$ instead of $k^3$. For more details on the universal IR scaling see Ref.~\cite{Cai:2019cdl}. The logarithmic running in the IR tail seems to be a characteristic of induced GWs in a radiation dominated universe \cite{Cai:2018dig,Yuan:2019wwo}. However, it is not present for general values of $w$ \cite{Domenech:2020kqm}.

Lastly, we consider the value $y\gg 1$, which corresponds to $k\gg k_{\rm peak}$. Assuming the same power-law ansatz but with $n<0$, the high frequency, or Ultra-Violet (UV), the tail is given by \cite{Atal:2021jyo}
\begin{align}\label{eq:UVtail}
\Omega^{\rm UV}_{\rm GW,*}(k)\propto
\begin{cases}
{\cal P}_{\cal R}^2(k) & {\rm for}\quad 0>n>-4\\[3pt]
\left(\frac{k}{k_{\rm peak}}\right)^{4}{\cal P}_{\cal R}(k)& {\rm for}\quad n<-4
\end{cases}
\,.
\end{align}
From Eq.~\eqref{eq:UVtail} we see that for sharply peaked primordial spectra the UV tail is quickly suppressed. In fact, for infinitely sharp primordial spectra (i.e. a Dirac delta), there is a sharp cut-off at $k=2k_{\rm peak}$ due to momentum conservation. Namely, one cannot induce tensor modes with a frequency higher than the source term. Nevertheless, if the primordial spectrum is not too steep, we may obtain information from the UV tail. Examples of this are the inference of the magnitude of local non-Gaussianities \cite{Atal:2021jyo} or the presence of a second phase of slow-roll inflation after the enhancement of fluctuations \cite{Balaji:2022dbi}. We also note that the $k^4$ scaling of the second case in Eq.~\eqref{eq:UVtail} is sensitive to the value of $c_s$. When $c_s\to 1$, it gradually transitions to a $k^2$ scaling \cite{Balaji:2022dbi}. One can see the power-law scalings (or the sharp cut-off) in the induced GW spectra of  Fig.~\ref{fig:examplespectra}. In the IR tail, all cases decay as $k^3\ln^2 k$. Instead, the UV tail is more sensitive to the shape of the primordial spectrum.

We end with a helpful rough estimate of the peak amplitude of the GW spectrum today from Eqs.~\eqref{eq:GWstoday}, \eqref{eq:IRtail} and \eqref{eq:UVtail}, disregarding the presence of the resonant peak. This is given by
\begin{align}\label{eq:estimateOmega}
\Omega^{\rm peak}_{\rm GW,0}h^2\approx 10^{-6}\,{\cal P}^2_{\cal R}(k_{\rm peak})\,,
\end{align}
which is good enough so that it also applies to a flat primordial spectrum (see Eq.~\eqref{eq:scaleinvariant}). From Eq.~\eqref{eq:estimateOmega} we see that the amplitude of the primordial spectrum of fluctuations measured by CMB observations, around $10^{-9}$ \cite{Planck:2018jri}, yields unobservable induced GWs. We need an enhancement of more than $4$ orders of magnitude to have an observable GW signal.

\subsection{Other extensions \label{sec:extensionsSGWB}}

\begin{figure}
\includegraphics[width=\columnwidth]{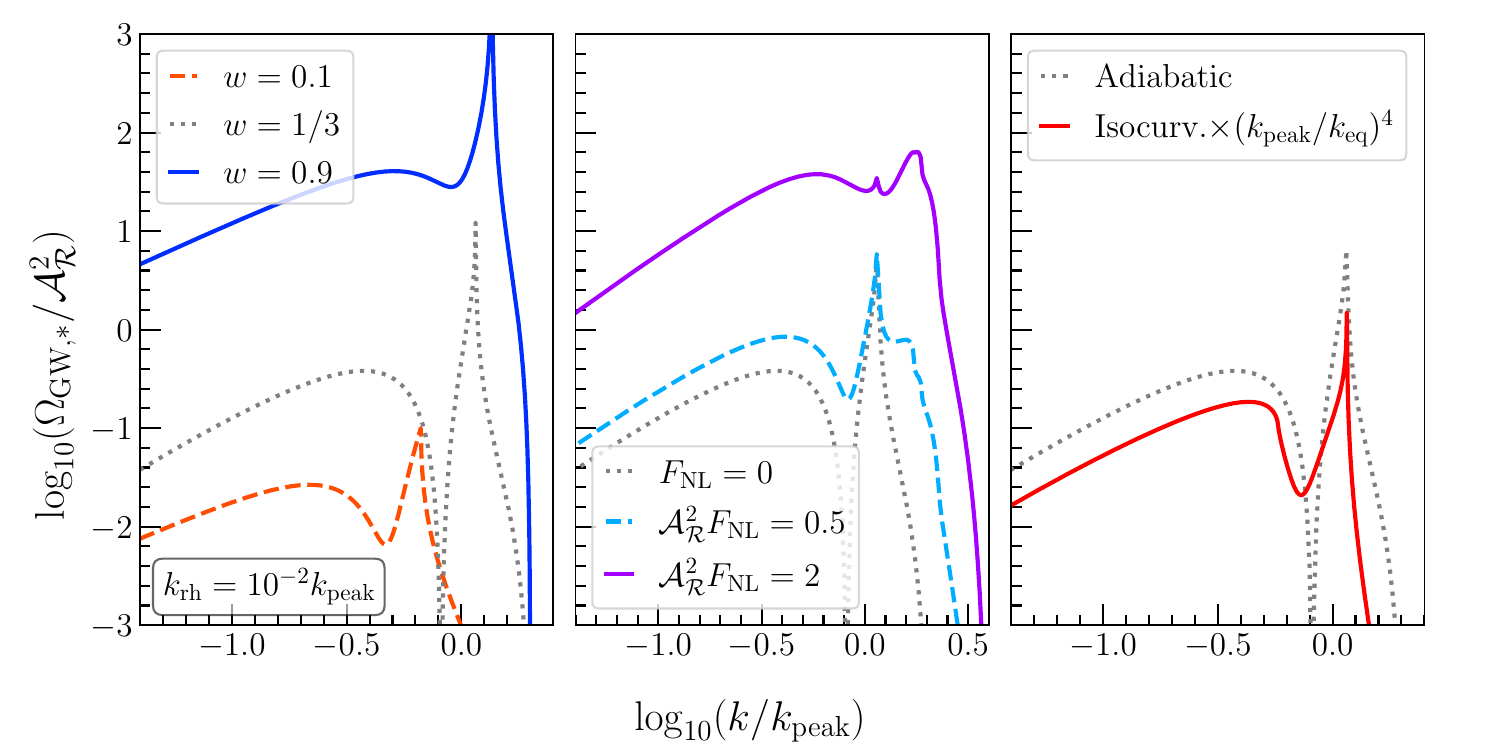}
\caption{ Induced GW spectrum for different extensions assuming a Dirac delta peak in the primordial spectrum of fluctuations. The radiation domination result is shown in grey lines. On the left panel, we show the effects of softer (orange) and stiffer (blue) equations of state $w$ from Ref.~\cite{Domenech:2019quo} (one may use SIGWfast \cite{Witkowski:2022mtg} for the finite width case). See how the stiffer the $w$, the higher the GW spectrum and the sharper the resonant peak (which moves to higher frequencies). On the middle panel, we show the effects of local non-Gaussianity from Ref.~\cite{Adshead:2021hnm} (data available at \cite{zachzenodo}). See how the effect of $F_{\rm NL}$ is to generate induced GW beyond the cut-off at $k=2k_{\rm peak}$ from Gaussian fluctuations. On the right panel, we compare the induced GW spectrum from adiabatic and isocurvature initial conditions from Ref.~\cite{Domenech:2021and}. Although the GW spectrum is qualitatively similar, they have a different shape of the resonant peak and the isocurvature-induced GWs have no destructive interference at $k=\sqrt{2}c_s k_{\rm peak}$.\label{fig:examplespectraextensions}}
\end{figure}

It is important to consider physics beyond the simplest model of Gaussian primordial fluctuations and a radiation-dominated universe. First, we have not probed the early universe at such small scales, so one should be as agnostic as possible. Second, there are several attractive models of inflation and reheating (and much more; see, e.g., Ref.~\cite{Allahverdi:2020bys} for a review) which predict departures such as non-Gaussianities of primordial fluctuation and a matter or kinetic dominated phase of reheating, respectively with $w\ll1$ (for an oscillating moduli field) and $w\approx 1$ (in quintessential inflation models). As we shall see, the induced GW spectrum depends on the initial conditions of primordial fluctuations (adiabatic/isocurvature and Gaussian/non-Gaussian) and the model of the primordial universe, e.g., the equation of state of the universe at the time of wave generation. Here we briefly mention below four different possibilities. We illustrate the effects of these possibilities on the induced GW spectrum in Fig.~\ref{fig:examplespectraextensions}.

\subsubsection{General equation of state of the very early universe \label{subsec:EoS}}
We know the universe had to be dominated by relativistic particles of the standard model around the time of BBN. More precisely, the radiation in the universe should have thermalised at $T\sim 4\,{\rm MeV}$ \cite{Kawasaki:1999na,Kawasaki:2000en,Hannestad:2004px,Hasegawa:2019jsa} for a successful BBN. We also have strong evidence for a period of cosmic inflation from CMB observations \cite{Planck:2018jri}. However, we do not know how or when the universe was re-heated. So, it is entirely plausible that there was a re-heating phase for $T>4\,{\rm MeV}$ with a different equation of state $w$ than radiation, e.g. a stiff fluid or a scalar field. If induced GWs (and PBHs) are generated during this $w$-dominated period, the shape of the induced GW spectral density will be modified. This is studied in detail in Refs.~\cite{Domenech:2019quo,Domenech:2020kqm}. The main changes are: $(i)$ an enhancement/suppression due to the different background expansion, $(ii)$ a different position of the resonant peak at $k_{\rm res}=2c_s k_{\rm peak}=2\sqrt{w}k_{\rm peak}$ and $(iii)$ a different power-law in the low-frequency tail going as $f^{3-2|b|}$ where $b=(1-3w)/(1+3w)$. We illustrate these effects on the left of Fig.~\ref{fig:examplespectraextensions}. In this way, induced GWs also probe the content of the primordial universe.

\subsubsection{Primordial non-Gaussianity and Trispectrum\label{subsec:non-Gaussianity}}

As we discussed around Eq.~\eqref{eq:h2pt}, the induced GW spectrum is determined by the 4-point function of curvature fluctuations. So, strictly speaking, induced GWs are also sensitive to any departure from Gaussian fluctuations. In general, this means that induced GWs probe the trispectrum of primordial fluctuations. However, if such fluctuations are generated during inflation, one expects perturbative departures from Gaussian fluctuations. And, therefore, the induced spectrum would only be slightly modified compared to the Gaussian case (see, e.g., Ref.~\cite{Garcia-Saenz:2022tzu} for a general treatment later applied to scale-invariant fluctuations). But, one could also assume a general ansatz for primordial non-Gaussianity and study its effects, regardless of the dynamics during inflation. The most studied case is local non-Gaussianity, which is parametrized as \cite{Komatsu:2003iq}
\begin{align}\label{eq:localNG}
\Phi(x)= \Phi_{\rm Gauss}(x)+ F_{\rm NL}(\Phi^2_{\rm Gauss}(x)-\langle\Phi^2_{\rm Gauss}(x)\rangle)\,,
\end{align}
where $\Phi_{\rm Gauss}$ is the Gaussian variable. See Ref.~\cite{Unal:2018yaa,Cai:2018dig,Atal:2021jyo} for early works and Ref.~\cite{Adshead:2021hnm} for a precise systematic approach. As shown in the middle panel of Fig.~\ref{fig:examplespectraextensions}, the local non-Gaussian contribution adds a resonant peak at $k=3c_sk_{\rm peak}$ and dominates over the Gaussian contribution for ${\cal A}^2_{\cal R}F_{\rm NL}>1$. Non-Gaussianity does not alter the predictions for the IR tail of the induced GW spectrum, though. Note that there is no reason to stop the ansatz \eqref{eq:localNG} at $\Phi^2$, especially if non-Gaussianities are large, as one could include higher powers of $\Phi$. See Ref.~\cite{Abe:2022xur} for a nice general treatment. As a curious possibility, we also mention that a parity-violating trispectrum could lead to chiral-induced GWs, though only as a small correction \cite{Garcia-Saenz:2023zue}.

\subsubsection{Isocurvature initial conditions \label{subsec:isocurvature}}

Observations of the CMB tell that primordial fluctuations on the largest scales are adiabatic \cite{Planck:2018jri}. This means that there is a time slice of uniform total energy density with only curvature perturbations but no relative number density fluctuations. But, at much smaller scales, this might not be the case. These initial conditions are called isocurvature \cite{Kodama:1986ud,Malik:2004tf}. Isocurvature fluctuations do not directly source induced GWs (see Ref.~\cite{Domenech:2021and} for the original work and Ref.~\cite{Domenech:2023jve} for a recent review). Instead, as the universe evolves, isocurvature fluctuations induce curvature fluctuations that later induce GWs. The price to pay is that the amplitude of induced GWs is suppressed by the energy density ratio of the subdominant field (responsible for isocurvature) with respect to the dominant field (responsible for curvature). For DM isocurvature fluctuations with a peaked isocurvature spectrum at $k_{\rm peak}$, the induced GW spectrum is suppressed by a factor $(k_{\rm eq}/k_{\rm peak})^{4}$, where $k_{\rm eq}\approx 0.01 \,{\rm Mpc}^{-1}$ is the comoving scale at the standard matter-radiation equality. Interestingly, the amplitude of the isocurvature spectrum can be very large, say ${\cal A}_{\cal S}\gg1$, and still be compatible with cosmological perturbation theory (up to the time fluctuations become non-linear and might form PBHs \cite{Passaglia:2021jla}). In this case, one has a detectable isocurvature-induced GW signal. Another interesting possibility is the GWs induced by the isocurvature from Poisson statistical fluctuations. This is a general signal when discrete objects or solitons exist, such as PBH, cosmic strings, oscillons, etcetera. This generic prediction has been dubbed ``Universal Gravitational Waves of Solitons'' in Refs.~\cite{Lozanov:2023aez,Lozanov:2023knf}. In Sec.~\ref{sec:reheatingSGWB}, we study in detail the GWs from PBH number density fluctuations.

\subsubsection{Induced GWs during inflation \label{subsec:GWsduringinflation}}

We mainly discussed the GWs induced by primordial fluctuations and the different extensions after inflation. However, induced GW can also be generated during inflation as a secondary effect from vacuum fluctuations  of spectator fields, such as scalar, vector, or tensor fields. The GW spectrum, in this case, is more model-dependent; therefore, we only cite the relevant literature. For scalar-induced GWs, e.g., in multifield models of inflation, we refer the reader to Refs.~\cite{Biagetti:2013kwa,Zhou:2020kkf,Fumagalli:2021mpc,Inomata:2021zel}. One typical case for vector field is axion inflation with Abelian or non-Abelian gauge fields, see Refs.~\cite{Barnaby:2011qe,Garcia-Bellido:2016dkw,Maleknejad:2016qjz}. For more references, we refer the reader to Komatsu's review on the topic \cite{Komatsu:2022nvu}. Two form fields can also source GWs at second order \cite{Fujita:2022ait}, though not at the linear order. Additional tensor fields can linearly source GWs \cite{Gorji:2023ziy}. For a review on linear GWs from metric vacuum fluctuations during inflation, see Ref.~\cite{Guzzetti:2016mkm}.

\section{GWs from the PBH dominated early universe \label{sec:reheatingSGWB}}

There is one scenario beyond the standard adiabatic primordial fluctuations which is interesting regarding the associated GW signals. This is the case when ultra-light PBHs dominate the very early universe and evaporate before BBN, reheating the universe via Hawking radiation \cite{Carr:1976zz,Lidsey:2001nj,Anantua:2008am}. Most relevant to secondary GWs is the fact that, as shown in Refs.~\cite{Inomata:2019ivs,Inomata:2019zqy,Inomata:2020lmk} (see also Ref.~\cite{Pearce:2023kxp}), an epoch of dust domination (such as a gas of PBHs) may enhance the production of induced GWs. Adding that, due to the discrete nature of PBHs, there are number density fluctuations from the Poisson noise in the PBH gas \cite{Papanikolaou:2020qtd}, one finds a unique induced GW signature of the PBH-dominated universe \cite{Papanikolaou:2020qtd,Domenech:2020ssp,Domenech:2021wkk}.  Below, we focus on the induced GW signatures following Ref.~\cite{Domenech:2020ssp}. The physics of BH evaporation is discussed in detail in Chapter 22.

\subsection{The PBH dominated universe \label{sec:PBHreheating}}

For simplicity, let us assume a monochromatic PBH mass function. In that case, there are two free parameters in the PBH reheating scenario: the PBH mass at formation $M_{\rm PBH,f}$ and the initial fraction $\beta$ of PBHs. The mass is proportional to the Hubble parameter at formation via Eq.~\eqref{eq:PBHmass}. The initial fraction is related to the \textit{mean} number density of PBHs per Hubble volume, that is
\begin{align}
  \beta=\frac{\rho_{\rm PBH,f}}{\rho_{\rm rad,\rm f}}=\frac{4\pi\gamma}{3H_{\rm f}^3}n_{\rm PBH,f}\,,
\end{align}
where we used that $\rho_{\rm PBH,f}=M_{\rm PBH,f}\times n_{\rm PBH,f}$ and $\rho_{\rm rad, \rm f}\approx 3H_{\rm f}^2M_{\rm pl}^2$. We can also compute the mean distance between PBHs from the mean number density. The comoving wavenumber associated with this distance is given by \cite{Papanikolaou:2020qtd}
\begin{align}\label{eq:kUV}
k_{\rm UV}=\left(\frac{4\pi n_{{\rm PBH},f}a_{\rm f}}{3}\right)^{1/3}=\beta^{1/3}\gamma^{-1/3}\,k_{\rm f}\,,
\end{align}
where $k_{\rm f}={\cal H}_{\rm f}$ is the comoving wavenumber that entered the horizon at formation.\footnote{Note that if such PBHs formed from the collapse of primordial fluctuations then $k_{\rm f}$ is equivalent to what we called $k_{\rm peak}$ in previous sections.} This means that PBHs are initially displaced over super-Hubble distances.

\begin{figure}
\includegraphics[width=\columnwidth]{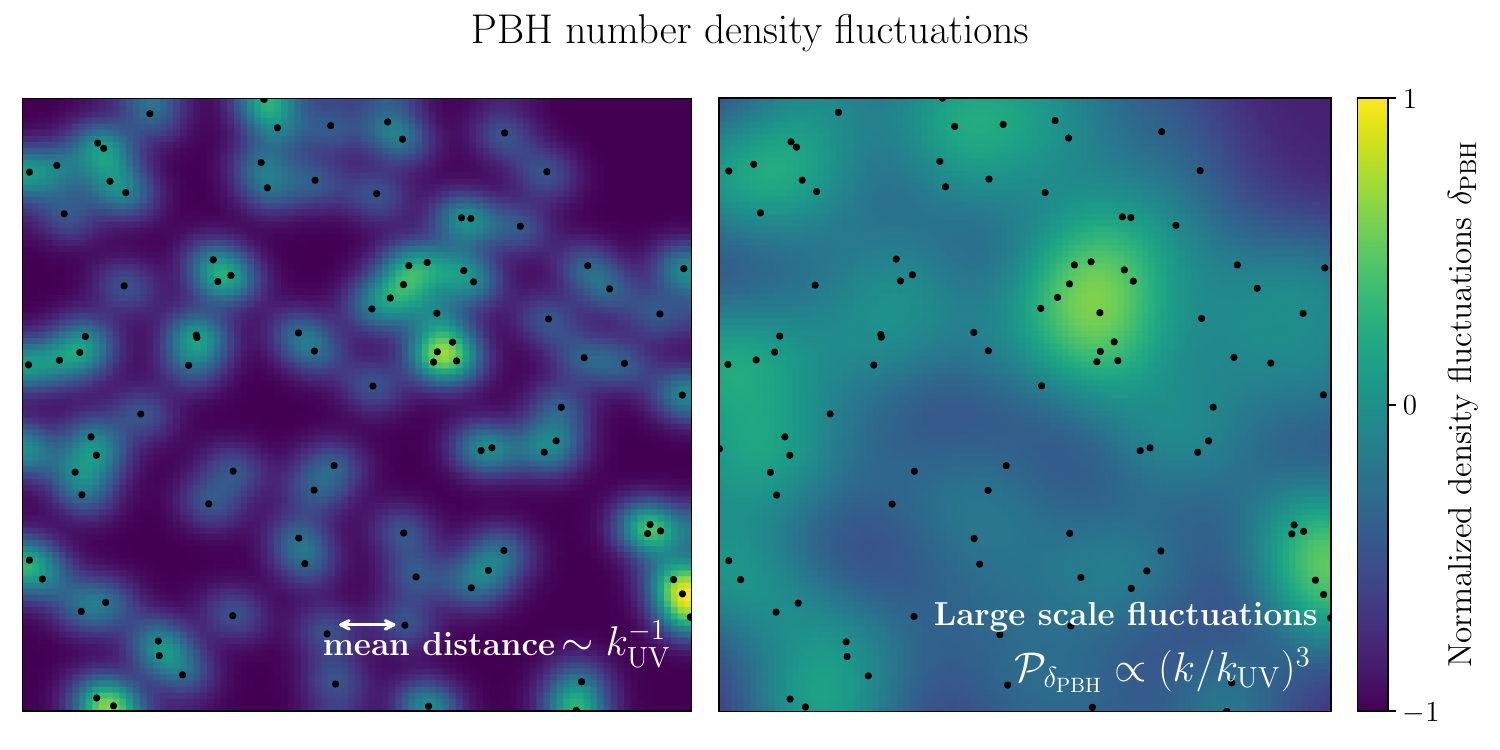}
\caption{PBH gas in the early universe. After formation, PBH are randomly distributed in space following a uniform distribution. This randomness leads to number density fluctuations. On the left panel, we show the kernel density estimate for the PBH number density in a fluid picture, smoothing on scales similar to the mean distance between PBHs. On the right panel, we show the kernel density estimate using a larger smoothing scale. See how number density fluctuations lead to large-scale density fluctuations of the PBH gas.\label{fig:illustrationPBHfluct}}
\end{figure}

The most interesting point is to note that, in fact, we have a gas of PBHs spread all over the universe. And, while it has on average a homogeneous distribution, there are statistical fluctuations. Namely, by chance, PBHs will be somewhere closer and somewhere sparser leading to number density fluctuations. Since PBHs form very rarely and, therefore, are mostly distributed uniformly in space, the distribution of PBHs follows Poisson statistics \cite{Papanikolaou:2020qtd}. The typical scale of the fluctuations is set by $k_{\rm UV}$ \eqref{eq:kUV}. The dimensionless power spectrum of PBH number density fluctuations at formation is given by \cite{Papanikolaou:2020qtd}
\begin{align}\label{eq:PS}
{\cal P}_{\delta_{\rm PBH,f}}(k)=\frac{2}{3\pi}\left(\frac{k}{k_{\rm UV}}\right)^3\Theta(k_{\rm UV}-k)\,,
\end{align}
where $\delta_{\rm PBH,f}=\delta n_{\rm PBH,f}/n_{\rm PBH,f}$ and $\Theta(x)$ is the Heaviside theta function. Note that the picture of a PBH gas is only valid if we have at least one PBH in a given volume. Thus, in the PBH gas approximation, we have a cut-off at the mean inter-PBH distance, namely at $k=k_{\rm UV}$. Hence, the notation ``UV'' to refer to the high wavenumber cut-off of the spectrum. We illustrate the PBH number density fluctuations in Fig.~\ref{fig:illustrationPBHfluct}.

After formation, PBHs evaporate via Hawking radiation. This leads to a cosmic time-dependent PBH mass given by \cite{Inomata:2020lmk}
\begin{align}\label{eq:MPBH(t)}
M_{\rm PBH}(t)\approx M_{{\rm PBH},\rm f}\left(1-\frac{t}{t_{\rm eva}}\right)^{1/3}\quad {\rm where}\quad 
t_{\rm eva}= \frac{M_{\rm PBH,f}^3}{M_{\rm pl}^4}\frac{160}{3.8\pi g_H(T_{\rm PBH})}\,.
\end{align}
The subscript ``eva'' refers to evaluation at evaporation. Because of PBH number density conservation, we have that $n_{\rm PBH}\propto a^{-3}$. This result together with Eq.~\eqref{eq:MPBH(t)} completely determines the time evolution of $\rho_{\rm PBH}$. Note that, during the time that Hawking evaporation is not effective, we have that $\rho_{\rm PBH}/\rho_{\rm r}\propto a$ and, eventually, PBHs might dominate the very early universe. 
The requirement that PBHs dominate the very early universe, which is equivalent to the requirement that PBHs evaporate after the Early PBH-radiation Equality (let us call this time ``eeq''), yields
\begin{align}\label{eq:betamin}
\beta>6\times10^{-10}\left(\frac{g_H(T_{\rm PBH,f})}{108}\right)^{1/2}
\left(\frac{\gamma}{0.2}\right)^{-1/2}\left(\frac{M_{{\rm PBH},\rm f}}{10^4{\rm g}}\right)^{-1}\,,
\end{align}
where $g_H(T_{\rm PBH})$ are the spin weighted degrees of freedom and $T_{\rm PBH,f}$ is the Hawking temperature of the PBH. For the standard model of particle physics, one has $g_H(T_{\rm PBH})\approx 108$. The temperature of the radiation filling the universe after PBH evaporation (which follows from saying that $H_{\rm eva}=2/(3t_{\rm eva})$, where the numerical factor is that of a matter-dominated universe) is given by
\begin{align}\label{eq:Teva}
T_{\rm eva}\approx 2.8\times 10^4\,{\rm GeV}\left(\frac{M_{{\rm PBH},\rm f}}{10^4{\rm g}}\right)^{-3/2}
\left(\frac{g_H(T_{\rm PBH, \rm f})}{108}\right)^{1/2}\left(\frac{g_\rho(T_{\rm eva})}{106.75}\right)^{-1/4}\,.
\end{align}
Imposing BBN bounds, namely that $T_{\rm eva}>4\,{\rm MeV}$ \cite{Kawasaki:1999na,Kawasaki:2000en,Hannestad:2004px,Hasegawa:2019jsa}, yields an upper bound on the PBH mass as
\begin{align}\label{eq:boundonmassBBN}
M_{{\rm PBH},f}<5\times 10^8{\rm g}\,.
\end{align}
The dynamics of the PBH reheating scenario are illustrated in Fig.~\ref{fig:examplePBHreheating}.

\begin{figure}
\includegraphics[width=\columnwidth]{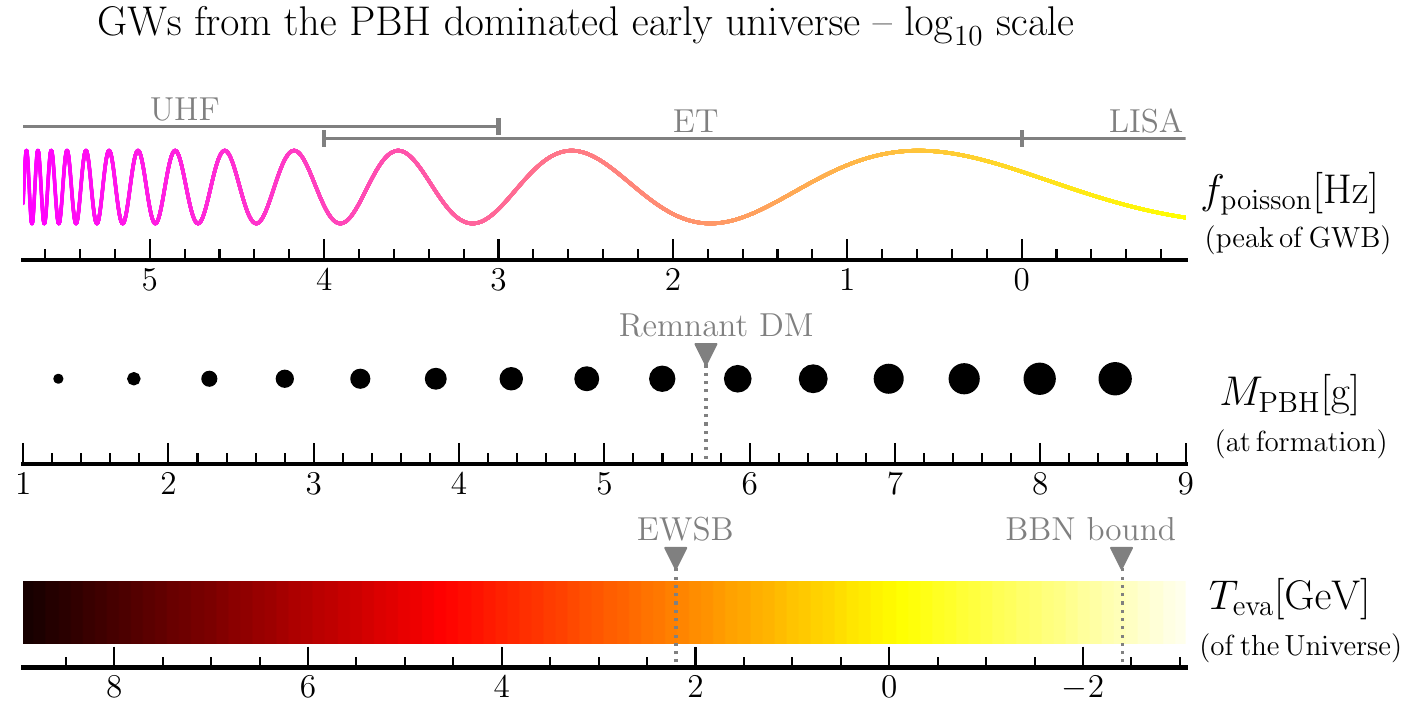}
\caption{Graphical summary of the GW signatures of the PBH-dominated universe. In the top panel, we show the frequency of the induced GWs associated with PBH reheating, $\log_{10}f_{\rm possion}$ \eqref{eq:fpoisson}. GW detectors in the relevant frequency range are UHF GW detectors, ET and LISA. In the middle panel, we show the corresponding PBH mass at formation, $\log_{10}M_{\rm PBH,f}$. We also show the PBH remnant scenario as DM \cite{MacGibbon:1987my}, which requires $M_{\rm PBH,f}\approx 5\times 10^5{\rm g}$ \cite{Domenech:2023mqk}. In the bottom panel, we give the temperature of the radiation filling the universe right after PBH evaporation $\log_{10}T_{\rm eva}$ \eqref{eq:Teva}. We show the BBN constraint $T\sim 4\,{\rm MeV}$ \cite{Kawasaki:1999na,Kawasaki:2000en,Hannestad:2004px,Hasegawa:2019jsa} as well as the temperature corresponding to ElectroWeak Symmetry Breaking (EWSB), which is relevant for leptogenesis/baryogenesis, see e.g. Ref.~\cite{Hooper:2020otu}. Interestingly, leptogenesis via Right-handed neutrinos requires $M_{\rm PBH,f}\lesssim 10\,{\rm g}$ \cite{Datta:2020bht,Bernal:2022pue}.\label{fig:illustrationPBHreheating}}
\end{figure}

It is also useful to compute the wavenumber corresponding to the Horizon size at evaporation (e.g. relative to the Hubble radius today). This is given by
\begin{align}\label{eq:keva}
k_{\rm eva}\approx 4.7\times 10^{11}{\rm Mpc}^{-1}\left(\frac{M_{{\rm PBH},\rm f}}{10^4{\rm g}}\right)^{-3/2}&\left(\frac{g_H(T_{\rm PBH})}{108}\right)^{1/2}\nonumber\\& \times\left(\frac{g_\rho(T_{\rm eva})}{106.75}\right)^{1/4}\left(\frac{g_{s}(T_{\rm eva})}{106.75}\right)^{-1/3}\,,
\end{align}
Since $T_{\rm eva}$ \eqref{eq:Teva} only depends on $M_{\rm PBH,f}$ and not $\beta$, the same applies to $k_{\rm eva}$. The frequency associated with the numerical factor in Eq.~\eqref{eq:keva}, namely $4.7\times 10^{11}{\rm Mpc}^{-1}$, is about $7.3\times 10^{-4}\,{\rm Hz}$. It turns out that, because of number density conservation (which implies $a\propto n_{\rm PBH}^{-1/3}$), $k_{\rm eva}$ computed in terms of $k_{\rm f}$ has the same $\beta$ dependence as $k_{\rm UV}$ \eqref{eq:kUV}. This implies that the ratio $k_{\rm UV}/k_{\rm eva}$ is independent of $\beta$. And, since $k_{\rm eva}$ from \eqref{eq:keva} depends on $M_{\rm PBH,f}$, the frequency associated with the cut-off of Poisson fluctuations (say $f_{\rm poisson}={k_{\rm UV}}/{2\pi}$) reads
\begin{align}\label{eq:fpoisson}
f_{\rm possion}&\approx 1.7\times 
10^{3}{\rm Hz}\left(\frac{M_{{\rm PBH},f}}{10^4{\rm g}}\right)^{-5/6}\left(\frac{g_H(T_{\rm PBH})}{108}\right)^{1/6}\left(\frac{g_\rho(T_{\rm eva})}{106.75}\right)^{1/4}
\left(\frac{g_{s}(T_{\rm eva})}{106.75}\right)^{-1/3}\,.
\end{align}
This frequency is where the induced GW spectrum from PBH Poisson fluctuations peaks. We show the parameter space for $f_{\rm poisson}$, $M_{\rm PBH,f}$ and $T_{\rm eva}$ in Fig.~\eqref{fig:illustrationPBHreheating}. See how GW detectors such as the LVK collaboration, Einstein Telescope and Cosmic Explorer can probe a wide range of masses of ultra-light PBHs if such PBHs dominate the universe.

\subsection{GWs from PBH number density fluctuations  \label{subsec:GWPBHdomination}}

As mentioned at the beginning of the section, PBHs are initially inhomogeneously distributed in space, according to Poisson statistics. The initial conditions of such number density fluctuations are isocurvature in nature \cite{Papanikolaou:2020qtd,Domenech:2020ssp}. This is because as each PBH forms, it leaves a ``hole'' in the original radiation fluid. Thus, by energy conservation, one must have in the fluid picture that $\delta\rho_{\rm PBH}+\delta\rho_{\rm rad}=0$ initially. This is nothing but isocurvature initial fluctuations. For instance, in the Newton gauge and on superhorizon scales, one has that $2\Phi\sim \delta\rho_{\rm tot}=\delta\rho_{\rm PBH}+\delta\rho_{\rm rad}=0$, i.e. there are no initial curvature fluctuations. But, as the universe evolves, and the mean PBH energy density starts to overtake that of the radiation fluid (see Fig.~\eqref{fig:illustrationPBHreheating}), there is a conversion from isocurvature to curvature fluctuations. The conversion is completed once PBHs dominate the universe. Then, if we call $S_i$ the initial isocurvature, one finds that the curvature fluctuations at PBH domination are given by $\Phi(k\ll k_{\rm eeq})=S_i/5$ and $\Phi(k\gg k_{\rm eeq})\approx \tfrac{3}{4}S_i(k/k_{\rm eeq})^{-2}$ (see e.g. Ref.~\cite{Kodama:1986ud}), where $k_{\rm eeq}=\sqrt{2}\,\beta^{2/3}\gamma^{1/3} k_{\rm UV}$ is the scale that enters the horizon at PBH-radiation equality. Scales with $k\gg k_{\rm eeq}$ are suppressed because they entered the horizon during the radiation-dominated epoch when the subhorizon curvature fluctuations decay. 

\begin{figure}
\includegraphics[width=\columnwidth]{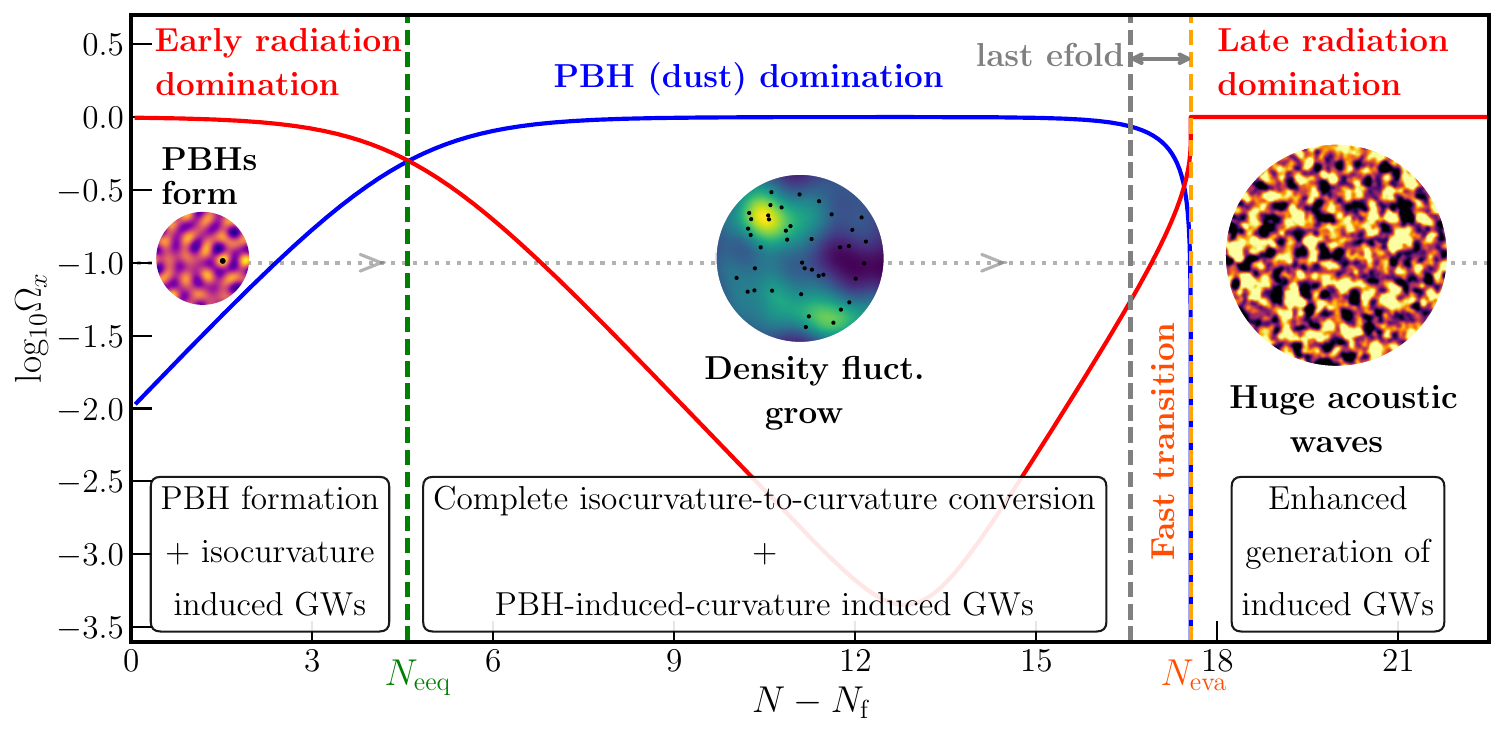}
\caption{Illustration of the universe's evolution in the PBH reheating scenario. We show $\Omega_{\rm PBH}=\rho_{\rm PBH}/\rho_{\rm tot}$ (with a blue line) and $\Omega_{\rm rad}=\rho_{\rm rad}/\rho_{\rm tot}$ (with a red line) in terms of efolds $N$ from PBH formation $N_{\rm f}$. PBHs start to dominate the universe at $N_{\rm eeq}$ and evaporate completely at $N_{\rm eva}$. During the early radiation domination, PBHs form, but the induced GW signal associated with the formation and PBH isocurvature are respectively of too high frequency or too low amplitude. After $N_{\rm eeq}$, PBH isocurvature has been converted into curvature fluctuations and induced GW are more efficiently produced (derived in Ref.~\cite{Papanikolaou:2020qtd}). Because the complete evaporation of PBHs is very fast, there is an enhanced production of induced GWs right after PBH evaporation \cite{Inomata:2020lmk} (estimated in Ref.~\cite{Domenech:2020ssp}).  \label{fig:examplePBHreheating}}
\end{figure}

\begin{figure}
\includegraphics[width=\columnwidth]{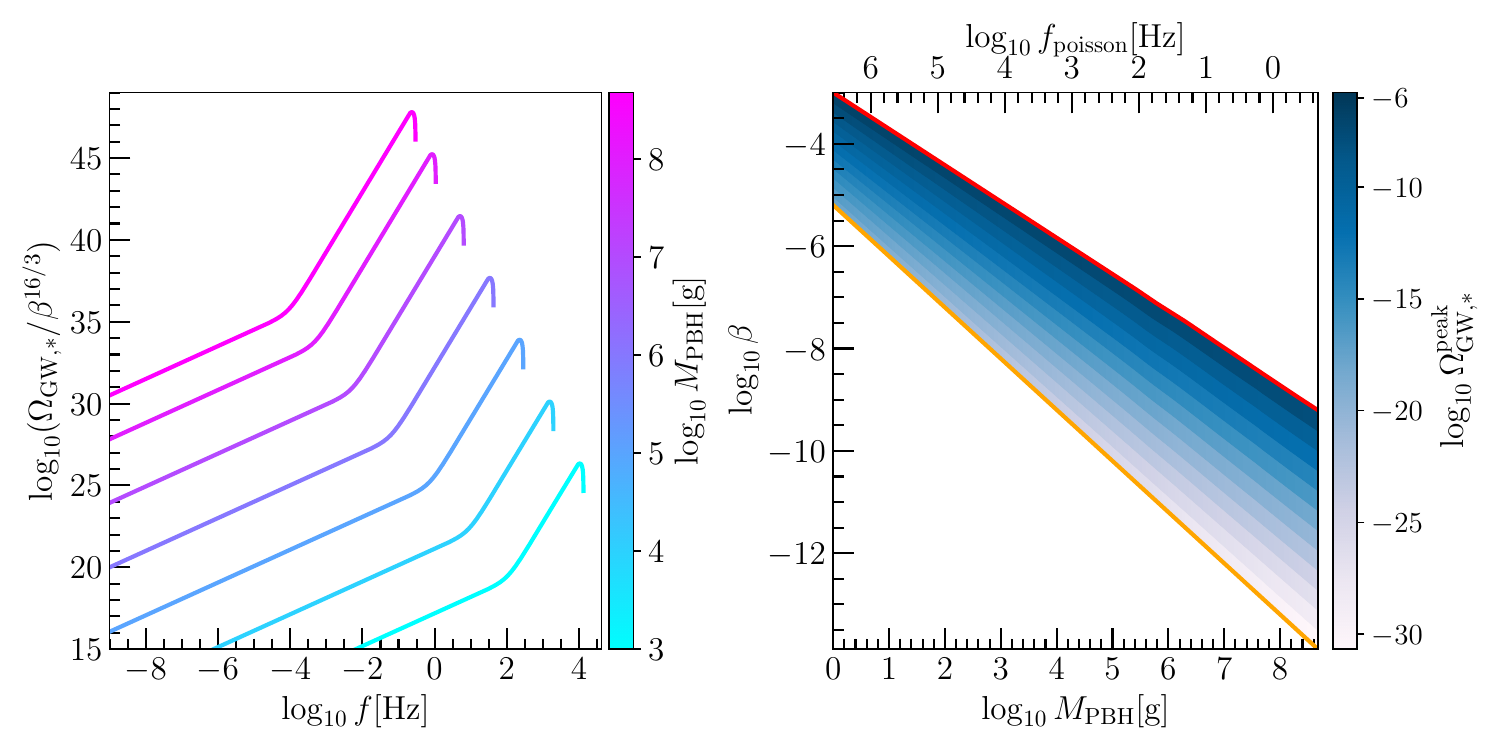}
\caption{On the left panel, we show the GW spectrum of GWs induced by PBH density fluctuations after PBH evaporation in terms of frequency for various values of $M_{\rm PBH,f}$ (shown in colour code) and normalized to $\beta^{16/3}$. The larger the mass, the higher the amplitude and the lower the peak frequency. On the right panel, we show the allowed parameter space of initial PBH fraction $\beta$ requiring PBH domination (lower bound \eqref{eq:betamin}, orange line) and using BBN constraints on the effective number of relativistic species (upper bound \eqref{eq:betamax}, red line). In colour code, we show the corresponding amplitude of the peak of the GW spectrum \eqref{eq:gwspeak}. \label{fig:examplespectrapoisson}}
\end{figure}

During the PBH-dominated phase, curvature fluctuations remain constant on all scales due to the zero pressure of the PBH gas. This also implies that PBH density fluctuations grow. During that period, there is a constant source of the induced GWs. Because of this, the induced GW spectrum for wavenumbers such that $k_{\rm eeq}<k<k_{\rm UV}$ turns out to be flat with an amplitude approximately given by \cite{Papanikolaou:2020qtd}
\begin{align}\label{eq:flatPBHdom}
\Omega^{\rm flat}_{\rm GW,PBH-dom}
\approx
2\times 10^{13} \left(\frac{ M_{\rm PBH,f} }{10^{4}\mathrm{g}}\right)^{4/3}\beta^{16/3}\,.
\end{align}
However, the biggest production of induced GWs happens right after PBH evaporate completely. The source of the enhancement is that by the time PBHs evaporate, the number of density fluctuations has grown very much. Then, the almost sudden transition from PBH to radiation domination (see in Fig.~\eqref{fig:examplePBHreheating} that the transition takes less than a quarter of efold) leads to large radiation density fluctuations (this is called the ``poltergeist'' mechanism in Ref.~\cite{Inomata:2020lmk}). Velocity fluctuations of the radiation fluid after evaporation are very large. From Eq.~\eqref{eq:vandPhi}, we find that after evaporation $v\sim \Phi'/{\cal H}\propto (k/k_{\rm eva})^2$. Taking into account that the transition is not instantaneous leads to a suppression factor $\sim (k/k_{\rm eva})^{-1/3}$ \cite{Inomata:2020lmk} (that comes from the fact that PBHs lose energy according to $M_{\rm PBH}\propto (1-t/t_{\rm eva})^{1/3}$). An approximation to the induced GW spectrum after PBH evaporation is given by \cite{Domenech:2020ssp,Domenech:2021wkk}
\begin{align}
\Omega_{\rm GW,eva}(k)\approx \Omega_{\rm GW,eva}^{\rm peak}\left(\frac{k}{k_{\rm UV}}\right)\Theta(k_{\rm UV}-k)\,,
\end{align}
where
\begin{align}\label{eq:gwspeak}
\Omega_{\rm GW,eva}
\approx10^{30}\beta^{16/3}\left(\frac{\gamma}{0.2}\right)^{8/3}\left(\frac{g_H(T_{\rm PBH})}{108}\right)^{-17/9}
\left(\frac{M_{{\rm PBH},\rm f}}{10^4{\rm g}}\right)^{34/9}\,.
\end{align}
Note that Eq.~\eqref{eq:gwspeak} is larger than Eq.~\eqref{eq:flatPBHdom} by a factor $M_{{\rm PBH},\rm f}^{22/9}$.  We show on the left of Fig.~\eqref{fig:examplespectrapoisson} the shape of the induced GW spectrum after PBH evaporation for different values of the PBH mass. On the right of Fig.~\eqref{fig:examplespectrapoisson}, we show the bounds on $\beta$ as a function of PBH mass and the corresponding amplitude of the induced GW spectrum peak \eqref{eq:gwspeak}.

It is important to note, though, that while the amplitude of Eq.~\eqref{eq:gwspeak} is very sensitive to the monochromatic mass function assumption (as a broad mass function leads to a more gradual transition to radiation domination), the amplitude of Eq.~\eqref{eq:flatPBHdom} is not as it only depends on the length PBH dominated phase. We can now use Eq.~\eqref{eq:gwspeak} to derive the current upper bounds on the initial fraction of PBHs using BBN bounds \cite{Cyburt:2004yc,Arbey:2021ysg} on the effective number of relativistic species (yielding $\Omega_{\rm GW,0}h^2\lesssim 10^{-6}$). This leads us to
\begin{align}\label{eq:betamax}
\beta<1.1\times 10^{-6}\left(\frac{\gamma}{0.2}\right)^{-1/2}\left(\frac{g_H(T_{\rm PBH})}{108}\right)^{17/48}
\left(\frac{g_*(T_{\rm eva})}{106.75}\right)^{1/16}\left(\frac{M_{{\rm PBH},\rm f}}{10^4{\rm g}}\right)^{-17/24}\,.
\end{align}

We end this section by mentioning shortcomings in the GW calculations of the PBH-dominated universe. First, although curvature fluctuations remain well inside the regime of cosmological perturbation theory, density fluctuations eventually become non-linear during PBH domination. For adiabatic primordial fluctuations, one may impose a cut-off to the primordial spectrum as in Ref.~\cite{Inomata:2020lmk} to ensure that all fluctuations are linear at all times. However, PBH density fluctuations are intrinsic to the model, so an arbitrary cut-off may not be as justified. More accurate predictions require more sophisticated numerical simulations. Second, the GW enhancement is highly sensitive to the monochromatic mass assumption. A broad mass function will lead to a slower transition to the late radiation-dominated regime. According to Ref.~\cite{Inomata:2020lmk}, the resulting induced GW spectrum is exponentially sensitive to the width of the mass function. It would be interesting to do a more detailed study.

\section{GW background from PBH binaries \label{sec:binariesSGWB}}

For completeness, we dedicate the last section of this chapter to the GW background from unresolved PBH binaries. We want to mention that this short section does not do justice to such an interesting topic. The interested reader is encouraged to read Ref.~\cite{Regimbau:2022mdu} for a recent review (see also Refs.~\cite{Regimbau:2011rp,Rosado:2011kv} for earlier reviews) and Refs.~\cite{Wang:2019kaf,Braglia:2021wwa} for applications to PBHs.

The GW background from PBH binaries has a different nature than the induced GW background. The former results from the PBH themselves as a collection of unresolved PBH mergers, while the latter is generated by the large-scale structure of fluctuations (very few of which collapse to PBHs). The first is the local, continuous, source of GWs (active until PBHs evaporate), while the second is only generated in the very early universe. The GW spectral density from unresolved PBH binaries is computed as the integrated energy flux in GWs per logarithmic frequency bin normalized to the critical energy density of the universe $\rho_c$ \cite{Zhu:2011bd,Braglia:2021wwa,Regimbau:2022mdu}. Schematically one has that \cite{Regimbau:2022mdu}
\begin{align}
	\Omega_{\rm GW}(f)=\frac{1}{\rho_c}fF_{\rm GW}(f)\,,
\end{align}
where $F_{\rm GW}$ is the integrated flux per frequency bin.\footnote{To be precise, $F_{\rm GW}$ also contains an integral over the direction of observation \cite{Regimbau:2022mdu}. However, we assume an isotropic background of binaries and consider the average GW energy density.} The integrated flux is nothing but an integral over time of the flux emitted per binary times the rate at which PBHs merge. The flux per binary is computed as the emitted energy in GWs per binary and per frequency bin, say ${dE_{\rm GW}}/{df}$, times the inverse area that the GWs expanded, namely $\tfrac{1}{4\pi r^2}\tfrac{dE_{\rm GW}}{df}$ where $r$ is the distance to the source. A nice analytical fit to ${dE_{\rm GW}}/{df}$, including the inspiral and merger phases, for non-precessing binaries is provided by Refs.~\cite{Ajith:2007kx,Ajith:2009bn} (they provide the template for the strain in frequency domain but the energy is proportional to $f^2$ times the strain squared). The rate at which the PBH binaries merge as a function of time (i.e. the number of mergers per unit time and unit volume) for PBHs is computed in Chapter 17. Here, we call it ${\cal R}_{\rm merge}(t)$. Using the cosmic redshift as a time variable,  i.e. $1+z=a(t)^{-1}$, the spectral density of the GW background from PBH binaries is given by \cite{Regimbau:2022mdu}
\begin{align}\label{eq:Omegabinaries}
\Omega^{\rm binaries}_{\rm GW}(f)=\frac{f}{3H_0^2M_{\rm pl}^2}\int_{0}^{z_{\rm max}}dz \frac{{\cal R}_{\rm merge}(z)}{(1+z)H(z)}\frac{dE_{\rm GW}}{df_s}\,,
\end{align}
where we used the notation $f_s$ to denote the frequency in the source frame, which is related to observed frequency by $f_s=f(1+z)$. If the GW background is composed by the merger of long-lived PBHs (i.e. $M_{\rm PBH,f}>10^{15}\,{\rm g}$), then only the nearby binaries (in the cosmological sense) contributes to the integral. In this case, we can take the Hubble parameter of the $\Lambda$CDM universe, i.e. $H(z)=H_0\sqrt{\Omega_{m,0}(1+z)^3+\Omega_{\Lambda,0}}$. The explicit expression for ${dE_{\rm GW}}/{df_s}$ for non-spinning PBHs (which is the expectation for PBH binaries at late times) reads \cite{Braglia:2021wwa,Regimbau:2022mdu}
\begin{align}\label{eq:dEdf}
\frac{dE_{\rm GW}}{df_s}=\frac{(G\pi)^{2/3}{\cal M}_c^{5/3}}{3}f_s^{-1/3}\times
\begin{cases}
(1-(1.44-2.68\eta) \nu^{2/3})^2&{\rm for}\,\, f_s<f_1\,,\\
\omega_mf_s (1-1.89\nu+1.66\nu^2)&{\rm for}\,\, f_1\leq f_s<f_2\,,\\
\omega_r f_s^{7/3}\frac{\nu_4^4}{4(\nu-\nu_2)^2+\nu_4^2}&{\rm for}\,\, f_2\leq f_s<f_3\,,
\end{cases}
\end{align}
where we defined $\nu_{n}\equiv(\pi G{\cal M} f_{n} )^{1/3}$, ${\cal M}=m_1 + m_2$ is the total mass, ${\cal M}_c= (m_1m_2)^{3/5}/{\cal M}^{1/5}$ is the chirp mass and $\eta = m_1 m_2/{\cal M}^2$. The numerical parameters in Eq.~\eqref{eq:dEdf} are given by
\begin{align}
\nu_1^3  &= 0.07+0.64\eta-0.06\eta^2-7.09\eta^3\,\,\,, \,\,\,
\nu_2^3  = 0.19+0.15\eta-0.02\eta^2+2.33\eta^3\,,\nonumber   \\
\nu_3^3 &= 0.32 -0.13\eta -0.27\eta^2 +4.92\eta^3\,\,\,, \,\,\,
\nu_4^3 = 0.09 -0.41\eta +1.83\eta^2-2.87\eta^3\,.
\end{align}
Lastly, upper integration limit in Eq.~\eqref{eq:Omegabinaries} is given by $z_{\rm max}=f_{3}/f-1$.

We end this section with some rough estimates for equal mass PBH binaries from Ref.~\cite{Domenech:2021odz} to develop a bit of intuition. The merger rate of equal PBH binaries evaluated today in its simplest form is given by \cite{Nakamura:1997sm,Bird:2016dcv,Sasaki:2016jop,Ali-Haimoud:2017rtz}
\begin{align}\label{eq:RR}
{\cal R}_{\rm merge}&\approx  4\times 10^6\,\text{Gpc}^{-3} \text{yr}^{-1} \frac{{\rm f}_{\rm PBH}^2}{\left({\rm f}_{\rm PBH}^2+\sigma_{\rm eq}^2\right)^{21/74}}  \left(\frac{M_{\rm PBH}}{M_\odot}\right)^{-32/37}\,,
\end{align}
where ${\rm f}_{\rm PBH}$ is the fraction of PBHs as dark matter and  $\sigma_{\rm eq}^2\approx 2.5\times 10^{-5}$. Putting some numbers and using that only the closest binaries, which are the loudest, contribute to the peak of the GW spectrum, we find that
\begin{align}\label{eq:GWbinaries}
\Omega^{\rm peak}_{\rm GW,binary}h^2\approx 1.6\times 10^{-8}\left(\frac{M_{\rm PBH,f}}{M_\odot}\right)^{5/37}\left(\frac{{\rm f}_{\rm PBH}}{0.01}\right)^{53/37}\left(1+0.25\left(\frac{{\rm f}_{\rm PBH}}{0.01}\right)^{-2}\right)^{-21/74}\,.
\end{align}
The peak of the GW spectrum is then roughly at the ISCO frequency, namely
\begin{align}\label{eq:fisco}
f_{\rm ISCO}\approx 2.2\,{\rm kHz}\left(\frac{M_{\rm PBH,f}}{M_\odot}\right)^{-1}\,.
\end{align}
This concludes our estimates for the GW background from unresolved PBH binaries.

\section{Final remarks \label{sec:remarksSGWB}}

The induced GW background is a decisive observable for the PBH scenario from large primordial fluctuations. In fact, it is generally expected that any mechanism that forms PBHs is always accompanied by a loud GW signal (as discussed in Sec.~\ref{sec:birdSGWB}). The absence of an induced GW background would leave any PBH scenario in doubt. But, most important to us, any hint (or even discovery) of PBHs in a given mass range provides us with strong motivation to look for the induced GW background (see Sec.~\ref{sec:curvatureSGWB}). From the GW background, we may also learn about the type of primordial density fluctuations (e.g. adiabatic/isocurvature and Gaussian/non-Gaussian) as well as the content of the very early universe (e.g. the equation of state $w$); see Sec.~\ref{sec:extensionsSGWB}. We may also be able to test whether PBHs once dominated the very early universe, reheating it via Hawking radiation. And, it might give us hints of PBH remnants as DM, as we saw in Sec.~\ref{sec:reheatingSGWB}. We may also be able to observe the GW background from the unresolved PBH binaries (see Sec.~\ref{sec:binariesSGWB}), providing further evidence to the PBH scenario. Besides the interesting observational consequences, induced GWs also motivate us to improve our theoretical understanding of GW energy density when dealing with second-order cosmological perturbation theory (as discussed in Sec.~\ref{sec:energySGWB}).

Interesting research directions that we have not discussed in this chapter are the study of statistical properties of the GW background (see, e.g. Ref.~\cite{Braglia:2022icu}) and GW background anisotropies (see, e.g., Ref.~\cite{Contaldi:2016koz,Bartolo:2019oiq,LISACosmologyWorkingGroup:2022kbp} for the general formulation and Ref.~\cite{Scelfo:2018sny,Wang:2021djr,Dimastrogiovanni:2022eir,Profumo:2023ybp,Bartolo:2019zvb,Li:2023qua,Yu:2023jrs} for applications to PBHs) as additional probes/evidence of the PBH scenario. We have also not discussed high-frequency GWs from Hawking evaporation of PBHs, which might be testable by its effects on the effective number of relativistic species at CMB and BBN \cite{Hooper:2019gtx,Masina:2020xhk,Masina:2021zpu,Arbey:2021ysg,Gehrman:2023esa,Ireland:2023avg}.  There is also the possibility to induce GWs from scalar-tensor and tensor-tensor mixings, see e.g. Refs.~\cite{Chang:2022vlv,Yu:2023lmo,Bari:2023rcw,Picard:2023sbz}.

We end this chapter by pointing out some directions worth exploring. First, one must examine the extensions beyond the simplest assumptions (radiation domination and Gaussian fluctuations) in more detail. This includes exploring motivated models that modify the expansion history of the early universe and derive more accurate kernels, studying how departures from General Relativity affect the induced GW spectrum, and, perhaps, exploring different types of non-Gaussianity of primordial fluctuations. Second, it would be very interesting to derive accurate predictions using numerical simulations for the GWs from the PBH-dominated early universe, as this is a unique way to probe the existence of PBHs that evaporated long ago. Lastly, it would be remarkable from the theoretical point of view to find a rigorous solution to the gauge ambiguity of induced GWs.

\section*{Acknowledgments} 
I would like to thank all of my collaborators from whom I learned a lot and without whom I would have probably never explored induced GWs to the extent I did. These are V.~Atal, S.~Balaji, P.~Bari, N.~Bartolo, E.~Dimastrogiovanni, M.~Fasiello, G.~Franciolini, A.~Malhotra, S.~Matarrese, A.~Naruko, C.~Lin, S.~Passaglia, S.~Patil, S.~Pi,  S.~Renaux-Petel, M.~Sasaki, V.~Takhistov, G.~Tasinato, T.~Vargas, A.~Wang, J.~Wang and  L.~Witkowski. I also thank J.~Fedrow, K.~Inomata, E.~Komatsu, T.~Papanikolaou, S.~Profumo, R.~Samanta, J.~Turner, C~.Unal, Z.~Weiner and S.~Young for many stimulating discussions

\bibliography{authorsample.bib} 

\end{document}